\documentclass[10pt,journal,onecolumn,final]{IEEEtran}

\usepackage[T1]{fontenc}

\ifCLASSINFOpdf
   \usepackage[pdftex]{graphicx}
   \graphicspath{{graphs/}, {bio_imgs/}}
  \DeclareGraphicsExtensions{.pdf,.jpg}
\else
   \usepackage[dvips]{graphicx}
   \graphicspath{{graphs/}, {bio_imgs/}}
   \DeclareGraphicsExtensions{.eps}
\fi

\usepackage{amsmath}

\usepackage{cases}
\usepackage[cmintegrals]{newtxmath}

\usepackage{cite}

\usepackage{tocloft}
\usepackage{setspace}
\usepackage{amsmath}
\usepackage{mathrsfs}
\usepackage{multicol}
\usepackage{amssymb}
\usepackage[utf8]{inputenc}
\usepackage[T1]{fontenc}
\usepackage[font={small}]{caption}
\usepackage{subfig}

\newtheorem{proposition}{Proposition}

\begin{document}
\title{D2D-Aware Device Caching in \\ MmWave-Cellular Networks}

%
%
%
\author{\IEEEauthorblockN{Nikolaos Giatsoglou, \emph{Member, IEEE}, Konstantinos Ntontin, \emph{Member, IEEE},
Elli Kartsakli, \emph{Senior Member, IEEE}, 
Angelos Antonopoulos, \emph{Senior Member, IEEE}, and Christos Verikoukis, \emph{Senior Member, IEEE}}
\thanks{
N. Giatsoglou and E. Kartsakli are with IQUADRAT Informatica S.L., Spain (e-mail: ngiatsoglou@iquadrat.com, ellik@iquadrat.com)

K. Ntontin is with the Department of Informatics and Telecommunications, University of Athens (UoA), Greece (e-mail: kntontin@di.uoa.gr).

A. Antonopoulos and C. Verikoukis are with the Telecommunications Technological Centre of Catalonia (CTTC/CERCA), Spain (e-mail: aantonopoulos@cttc.es,cveri@cttc.es)
}
}


\maketitle

\begin{abstract}
In this paper, we propose a novel policy for device caching that facilitates popular content exchange through high-rate device-to-device (D2D) millimeter-wave (mmWave) communication. The D2D-aware caching (DAC) policy splits the cacheable content into two content groups and distributes it randomly to the user equipment devices (UEs), with the goal to enable D2D connections. By exploiting the high bandwidth availability and the directionality of mmWaves, we ensure high rates for the D2D transmissions, while mitigating the co-channel interference that limits the throughput gains of D2D communication in the sub-6 GHz bands. Furthermore, based on a stochastic-geometry modeling of the network topology, we analytically derive the offloading gain that is achieved by the proposed policy and the distribution of the content retrieval delay considering both half- and full-duplex mode for the D2D communication. The accuracy of the proposed analytical framework is validated through Monte-Carlo simulations. In addition, for a wide range of a content popularity indicator the results show that the proposed policy achieves higher offloading and lower content-retrieval delays than existing state-of-the-art approaches.
\end{abstract}

\begin{IEEEkeywords}
caching policies, Zipf popularity model, stochastic-geometry, wireless full-duplex communication.
\end{IEEEkeywords}

%
\IEEEpeerreviewmaketitle

\section{Introduction}

\subsection{Background}

\IEEEPARstart{O}{ver} the last few years, the proliferation of mobile devices connected to the Internet, such as smartphones and tablets, has led to an unprecedented increase in wireless traffic that is expected to grow with an annual rate of 53\% until 2020\cite{general:cisco_report}. To satisfy this growth, a goal has been set for the 5th generation (5G) of mobile networks to improve the capacity of current networks by a factor of 1000\cite{general:andrews_whatwill5gbe}. While traditional approaches improve the area spectral efficiency of the network through, e.g., cell densification, transmission in the millimeter-wave (mmWave) band, and massive MIMO\cite{general:andrews_whatwill5gbe}, studies have highlighted the repetitive pattern of user content requests\cite{cacheability:http, Femtocaching_and_D2D}, suggesting more efficient ways to serve them.

With \textit{proactive caching}, popular content is stored inside the network during off-peak hours (e.g., at night), so that it can be served locally during peak hours\cite{general:proactive_caching}. Two methods are distinguished in the literature: i) \textit{edge caching}\cite{edgecaching:femtocaching} when the content is stored at helper nodes, such as small-cell base stations (BSs), and ii) \textit{device caching}\cite{devcaching:molisch_scalinglaw} when the content is stored at the user equipment devices (UEs). While edge caching alleviates the backhaul constraint of the small-cells by reducing the transmissions from the core network, device caching offloads the BSs by reducing the cellular transmissions, which increases the rates of the active cellular UEs and reduces the dynamic energy consumption of the BSs\cite{general:bs_power_consumption}. The UEs also experience lower delays since the cached content is served instantaneously or through D2D communication from the local device caches.

The benefits of device caching in the offloading and the throughput performance have been demonstrated in \cite{devcaching:molisch_scalinglaw, devcaching:molisch_throughput_outage_tradeoff, devcaching:molisch_clustering, devcaching:local_global_gains, devcaching:d2d_optimization, devcaching:maximized_traffic_offloading}. In \cite{devcaching:molisch_scalinglaw}, the spectrum efficiency of a network of D2D UEs that cache and exchange content from a content library, is shown to scale linearly with the network size, provided that their content requests are sufficiently redundant. In \cite{devcaching:molisch_throughput_outage_tradeoff}, the previous result is extended to the UE throughput, which, allowing for a small probability of outage, is shown to scale proportionally with the UE cache size, provided that the aggregate memory of the UE caches is larger than the library size. To achieve these scaling laws, the impact of the D2D interference must be addressed by optimally adjusting the D2D transmission range to the UE density. In \cite{devcaching:molisch_clustering}, a cluster-based approach is proposed to address the D2D interference where the D2D links inside a cluster are scheduled with time division multiple access (TDMA). The results corroborate the scaling of the spectrum efficiency that was derived in \cite{devcaching:molisch_scalinglaw}. In \cite{devcaching:local_global_gains}, a mathematical framework based on stochastic geometry is proposed to analyze the cluster-based TDMA scheme, and the trade-off between the cluster density, the local offloading from inside the cluster, and the global offloading from the whole network is demonstrated through extensive simulations. In \cite{devcaching:d2d_optimization}, the system throughput is maximized by jointly optimizing the D2D link scheduling and the power allocation, while in \cite{devcaching:maximized_traffic_offloading}, the offloading is maximized by an interference-aware reactive caching mechanism.

Although the aforementioned works show positive results for device caching, elaborate scheduling and power allocation schemes are required to mitigate the D2D interference, which limit the UE throughput and increase the system complexity. The high impact of the D2D interference is attributed to the omni-directional transmission patterns that are commonly employed in the sub-6 GHz bands. While directionality could naturally mitigate the D2D interference and alleviate the need for coordination, it requires a large number of antennas, whose size is not practical in the microwave bands. In contrast, the mmWave bands allow the employment of antenna arrays in hand-held UE devices due to their small wavelength. Combined with the availability of bandwidth and their prominence in future cellular communications\cite{general:andrews_whatwill5gbe}, the mmWave bands are an attractive solution for D2D communication \cite{d2dmmaves:enabling_article, d2dmmwaves:exploiting_access_backhaul}.

The performance of the mmWave bands in wireless communication has been investigated in the literature for both outdoor and indoor environments, especially for the frequencies of 28 and 73 GHz that exhibit small atmospheric absorption  \cite{mmwaves:rappaport_channel_modeling, mmwaves:rappaport_indoor_measurements}. According to these works, the coverage probability and the average rate can be enhanced with dense mmWave deployments when highly-directional antennas are employed at both the BSs and the UEs. MmWave systems further tend to be noise-limited due to the high bandwidth and the directionality of transmission\cite{mmwaves:andrews_kulkarni_rate_trends_for_blockage_param_values}. Recently, several works have conducted system-level analyses of mmWave networks with stochastic geometry\cite{mmwaves:heath_bai_coverage_and_rate, mmwaves:heath_bai_blockage_model_analysis, mmwaves:heath_bai_coverage_and_capacity}, where the positions of the BSs and the UEs are modeled according to homogeneous Poisson point processes (PPPs)\cite{stochgeom:haenggi_book}. This modeling has gained recognition due to its tractability\cite{stochgeom:andrews_tractable_classic}. 
\subsection{Motivation and Contribution}

Based on the above, it is seen that device caching can significantly enhance the offloading and the delay performance of the cellular network, especially when the UEs exchange cached content through D2D communication. On the other hand, the D2D interference poses a challenge in conventional microwave deployments due to the omni-directional pattern of transmission. While directionality is difficult to achieve in the sub-6 GHz band for hand-held devices, it is practical in the mmWave frequencies due to the small size of the antennas. The high availability of bandwidth and the prominence of the mmWave bands in future cellular networks have further motivated us to consider mmWave D2D communication in a device caching application. To the best of our knowledge, this combination has only been considered in \cite{devcaching:molisch_tutorial}, which adopts the cluster-based TDMA approach for the coordination of the D2D links and does not exploit the directionality of mmWaves to further increase the D2D frequency reuse.

In this context, the contributions of our work are summarized as follows:

\begin{itemize}

\item We propose a novel D2D-aware caching (DAC) policy for device caching that facilitates the content exchange between the paired UEs through mmWave D2D communication and exploits the directionality of the mmWave band to increase the frequency reuse among the D2D links. In addition, we consider a half-duplex (HD) and a full-duplex (FD) version of the DAC policy when  simultaneous requests occur inside a D2D pair.

\item We evaluate the performance of the proposed policy in terms of an offloading metric and the distribution of the content retrieval delay, based on a stochastic geometry framework for the positions of the BSs and the UEs.

\item We compare our proposal with the state-of-the-art most-popular content (MPC) policy through analysis and simulation, which shows that our policy improves the offloading metric and the 90-th percentile of the delay when the availability of  paired UEs is sufficiently high and the content popularity distribution is not excessively peaked.
\end{itemize}

The rest of the paper is organized as follows. In Section \ref{section:caching_model_and_policies}, we present the proposed DAC and the state-of-the-art MPC policy. In Section \ref{section:system_model}, we present the system model. In Section \ref{section:offloading_analysis} and Section \ref{section:performance_analysis}, we characterize the performance of the two policies in terms of the offloading factor and the content retrieval delay respectively. In Section \ref{section:results}, we compare analytically and through simulations the performance of the caching policies. Finally, Section \ref{section:conclusion} concludes the paper.

\section{Background and Proposed Caching Policy}
\label{section:caching_model_and_policies}
In this section, based on a widely considered model for the UE requests, we present the state-of-the-art MPC policy and the proposed DAC policy. 

\subsection{UE Request Model}
We assume that the UEs request content from a library of $L$ files of equal size $\sigma_{file}$\cite{edgecaching:femtocaching} and that their requests follow the Zipf distribution. According to this model, after ranking the files with decreasing popularity, the probability $q_i$ of a UE requesting the $i$-th ranked file is given by
\begin{equation}
q_i = \frac{i^{-\xi}}{\sum_{j=1}^L j^{-\xi}}, \mbox{  } 1 \leq i \leq L, \mbox{ } \xi \geq 0,
\end{equation}
where $\xi$ is the popularity exponent of the Zipf distribution. This parameter characterizes the skewness of the popularity distribution and depends on the content\footnote{Please note that the terms \textit{file} and \textit{content} are used interchangeably in the following.} type, (e.g., webpages, video, audio, etc.)\cite{cacheability:evidence_implications, cacheability:viewedge}. 

\subsection{State-of-the-Art MPC Policy}
In device caching, every UE retains a cache of $K$ files, where $K<<L$, so that when a cached content is requested, it is retrieved locally with negligible delay instead of a cellular transmission. This event is called a \textit{cache hit} and its probability is called the \textit{hit probability}, denoted by $h$ and given by
\begin{equation}
h = \sum_{i \in \mathcal{C}} q_i \label{eq:hitprob_def},
\end{equation}
where $\mathcal{C}$ represents the cached contents of a UE, as determined by the caching policy. The MPC policy is a widely considered caching scheme\cite{devcaching:molisch_seminal_conf, edgecaching:giovanidis_optimal_geographic_random_policy, devcaching:molisch_clustering} that stores the $K$ most popular contents from the library of $L$ files in every UE, resulting in the maximum hit probability, given by
\begin{equation}
h_{mpc}= \sum_{i=1}^K q_i = \frac{\sum_{i=1}^K i^{-\xi}}{\sum_{j=1}^{L} j^{-\xi}}.
\label{eq:hmpc}
\end{equation}

\subsection{Proposed DAC Policy}
Although the MPC policy maximizes the hit probability, it precludes content exchange among the UEs since all of them store the same files. In contrast, a policy that diversifies the content among the UEs enables content exchange through D2D communication, resulting in higher offloading. Furthermore, thanks to the high D2D rate and the enhancement in the cellular rate due to the offloading, the considered policy may also improve the content retrieval delay, despite its lower hit probability compared with the MPC policy.

Based on this intuition, in the proposed DAC policy, the $2 K$ most popular contents of the library of $L$ files are partitioned into two non-overlapping groups of $K$ files, denoted by groups A and B, and are distributed randomly to the UEs, which are characterized as UEs A and B respectively. When a UE A is close to a UE B, the network may pair them to enable content exchange through D2D communication. Denoting by $h_A$ and $h_B$ the hit probabilities of the two UE types, three possibilities exist when a paired UE A requests content:
\begin{itemize}
\item the content is retrieved with probability $h_A$ through a cache hit from the local cache of UE A.
\item the content is retrieved with probability $h_B$ through a D2D transmission from the cache of the peer UE B.
\item the content is retrieved with probability $1-h_A-h_B$ through a cellular transmission from the associated BS of UE A.
\end{itemize}
The above cases are defined accordingly for a paired UE B. In Proposition 1 that follows, we formally prove that the probability of content exchange for both paired UEs are maximized with the content assignment of the DAC policy.

\begin{proposition}
Denoting by $\mathcal{C}_A$ and $\mathcal{C}_B$ the caches of UE A and B inside a D2D pair, and by $e_A$ and $e_B$ their probabilities of content exchange, $e_A$ and $e_B$ are maximized when $\mathcal{C}_A$ and $\mathcal{C}_B$ form a non-overlapping partition of the $2K$ most popular contents, i.e., $\mathcal{C}_A \cup \mathcal{C}_B = \{i \in \mathbb{N}: 1 \leq i \leq 2 K \}$ and $\mathcal{C}_A \cap \mathcal{C}_B = \emptyset$, in the sense that no other content assignment to $\mathcal{C}_A$ and $\mathcal{C}_B$ can \textit{simultaneously} increase $e_A$ and $e_B$. 
\end{proposition}

\begin{IEEEproof}
See Appendix \ref{appendix:proposition_1}.
\end{IEEEproof}

When the paired UEs store non-overlapping content, their hit probabilities coincide with their content exchange probabilities, i.e., $e_A=h_B$ and $e_B=h_A$, hence, the DAC policy also maximizes $h_A$ and $h_B$ over all possible $2 K$ partitions in the sense of Proposition 1\footnote{Please note that $h_A$ and $h_B$ are still lower than $h_{mpc}$, since the MPC policy is not based on partitions.}. The $2K$ most popular contents can be further partitioned in multiple ways, but one that equalizes $h_A$ and $h_B$ is chosen for fairness considerations. Although exact equalization is not possible due to the discrete nature of the Zipf distribution, the partition that minimizes the difference $|h_A-h_B|$ can be found. Considering that this difference is expected to be negligible for sufficiently high values of $K$, $h_A$ and $h_B$ can be expressed as
\begin{equation}
h_A \approx h_B \approx h_{dac} = \frac{1}{2} \sum_{i=1}^{2K} q_i.
\end{equation}
Finally, since two paired UEs may want to simultaneously exchange content, with probability $h_{dac}^2$, we consider two cases for the DAC policy: i) an HD version, denoted by HD-DAC, where the UEs exchange contents with two sequential HD transmissions, and ii) an FD version, denoted by FD-DAC, where the UEs exchange contents simultaneously with one FD transmission. Although the FD-DAC policy increases the frequency reuse of the D2D transmissions compared with the HD-DAC policy, it also introduces self-interference (SI) at the UEs that operate in FD mode\cite{fd:survey} and increases the D2D co-channel interference. It therefore raises interesting questions regarding the impact of FD communication on the rate performance, especially in a mmWave system where the co-channel interference is naturally mitigated by the directionality.

\section{System Model}\label{section:system_model}
In this section, we present the network model, the mmWave channel model, the FD operation of the UEs, and the resource allocation scheme for the cellular and the D2D transmissions.

\subsection{Network Model}\label{sys:network}
\begin{figure}[!t]
\centering
\includegraphics[width=2.3in]{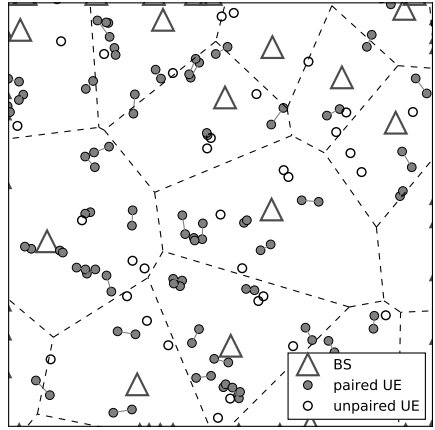}
\caption{A network snapshot in a rectangle of dimensions 300 m x 300 m consisting of BSs (triangles) and UEs (circles). The paired UEs are shown connected with a solid line.}
\label{fig:network_model}
\end{figure}

We consider a cellular network where a fraction of the UEs are paired, as shown in the snapshot of Fig. \ref{fig:network_model}. We assume that the BSs are distributed on the plane according to a homogeneous PPP $\Phi_{bs}$ of intensity $\lambda_{bs}$, while the UEs are distributed according to three homogeneous PPPs: the PPP $\Phi_u$ with intensity $\lambda_{u}$ representing the unpaired UEs, and the PPPs $\Phi_p^{(1)}$ and $\Phi_p^{(2)}$ with the same intensity $\lambda_{p}$ representing the paired UEs. We assume that $\Phi_u$ is independent of $\Phi_p^{(1)}$ and $\Phi_p^{(2)}$, while $\Phi_p^{(1)}$ and $\Phi_p^{(2)}$ are dependent due to the correlation introduced by the D2D pairings. Specifically, for every UE of $\Phi_{p}^{(1)}$, a D2D peer exists in $\Phi_{p}^{(2)}$ that is uniformly distributed inside a disk of radius $r_{d2d}^{max}$, or, equivalently, at a distance $r_{d2d}$ and an angle $\phi_{d2d}$ that are distributed according to the probability density functions (PDFs) $f_{r_{d2d}}(r)$ and $f_{\phi_{d2d}}(\phi)$, given by

\begin{subequations}
\begin{equation}
 f_{r_{d2d}}(r) = \frac{2 r}{(r_{d2d}^{max})^2}, \mbox{ } 0<r< r_{d2d}^{max},
\label{eq:d2d_displacement}
\end{equation}    
\begin{equation}
f_{\phi_{d2d}}(\phi) = \frac{1}{2 \pi},\mbox{ }   0 \leq \phi < 2\pi.
\end{equation}
\end{subequations}

We assume that the D2D pairings arise when content exchange is possible, based on the cached files of the UEs. In the DAC policy, the BSs distribute the content groups A and B independently and with probability 1/2 to their associated UEs, and a fraction $\delta$ of them, which are located within distance $r_{d2d}^{max}$, are paired. Defining the aggregate process of the UEs $\Phi_{ue}$ as
\begin{equation}
\Phi_{ue} \triangleq \Phi_{u} \cup \Phi_{p}^{(1)} \cup \Phi_{p}^{(2)},
\end{equation}
and its intensity $\lambda_{ue}$ as\footnote{Please note that $\Phi_{ue}$ is not a PPP due to the correlation introduced by the processes of the paired UEs, $\Phi_
{p}^{1}$ and $\Phi_{p}^{2}$. Nevertheless, its intensity can be defined as the average number of UEs per unit area.}
\begin{equation}
\lambda_{ue} = \lambda_u+2 \lambda_p,
\end{equation}
the ratio $\delta$ of the paired UEs is given by
\begin{equation}
\delta = \frac{2 \lambda_p}{\lambda_{ue}} = \frac{2 \lambda_p}{\lambda_u+2 \lambda_p}.
\end{equation}

Regarding the UE association, we assume that all the UEs are associated with their closest BS\footnote{Although different association criteria could have been considered, e.g., based on the maximum received power, the comparison of the two caching policies is not expected to be affected. Hence, we consider the closest BS association scheme due to its analytical tractability.}, in which case the cells coincide with the Voronoi regions generated by $\Phi_{bs}$. Denoting by $A_{cell}$ the area of a typical Voronoi cell, the equivalent cell radius $r_{cell}$ is defined as
\begin{equation}\label{def:rcell}
r_{cell} \triangleq \sqrt{ \frac{\mathbb{E} [A_{cell}]}{\pi}}=\frac{1}{\sqrt{\pi \lambda_{bs}}},
\end{equation}
and the association distance $r$ of a UE to its closest BS is distributed according to the PDF $f_r(r)$, given by\cite{stochgeom:andrews_tractable_classic}
\begin{gather}
f_{r_{}}(r)= \frac{2r}{r_{cell}^2} e^{-\left(\frac{r}{r_{cell}}\right)^2} =  2 \lambda_{bs}\pi r e^{-\lambda_{bs} \pi r^2}, \mbox{ } r>0.
\label{def:rassoc_pdf}
\end{gather}

\subsection{Channel Model}\label{sys:phy}
Regarding the channel model, we assume that the BSs and the UEs transmit with constant power, which is denoted by $P_{bs}$ and $P_{ue}$ respectively, and consider transmission at the mmWave carrier frequency $f_c$ with wavelength $\bar{\lambda}_c$ for both the cellular and the D2D communication through directional antennas employed at both the BSs and the UEs. The antenna gains are modeled according to the sectorized antenna model\cite{mmwaves:andrews_for_sectorized_model}, which assumes constant mainlobe and sidelobe gains, given by
\begin{subnumcases}{ G_{i}(\theta) =}
G_{i}^{max} &
if $|\theta| \leq \Delta\theta_{i}$, \\
G_{i}^{min} & 
if $|\theta| > \Delta\theta_{i}$,
\end{subnumcases}
where $\Delta\theta$ is the antenna beamwidth, $\theta$ is the angle deviation from the antenna boresight, and $i \in \left\lbrace bs, ue \right\rbrace$.

Because the mmWave frequencies are subject to blockage effects, which become more pronounced as the transmission distance increases \cite{mmwaves:rappaport_channel_modeling}, the line-of-sight (LOS) state of the mmWave links is explicitly modeled. We consider the \textit{exponential model}  \cite{mmwaves:heath_bai_blockage_model_analysis,mmwaves:rappaport_channel_modeling}, according to which a link of distance $r$ is LOS with probability $\mbox{P}_{los}(r)$ or non-LOS (NLOS) with probability $1-\mbox{P}_{los}(r)$, where $\mbox{P}_{los}(r)$ is given by
\begin{equation}
\mbox{P}_{los}(r) =e^{-\frac{r}{r_{los}}}.
\label{eq:plos}
\end{equation}
The parameter $r_{los}$ is the average LOS radius, which depends on the size and the density of the blockages \cite{mmwaves:heath_bai_blockage_model_analysis}. We further assume that the pathloss coefficients of a LOS and a NLOS link are $a_{L}$ and $a_{N}$ respectively, the LOS states of different links are independent, and the shadowing is incorporated into the LOS model \cite{mmwaves:heath_feasibility_spectrum_licenses_for_no_lognormal}. Finally, we assume Rayleigh fast fading where the channel power gain, denoted by $\eta$, is exponentially distributed, i.e., $\eta \! \sim \! Exp(1)$.

\subsection{FD-Operation Principle}
When a UE operates in FD mode, it receives SI by its own transmission. The SI signal comprises a direct LOS component, which can be substantially mitigated with proper SI cancellation techniques, and a reflected component, which is subject to multi-path fading. Due to the lack of measurements regarding the impact of the aforementioned components in  FD mmWave transceivers, we model the SI channel as Rayleigh \cite{fd:si_model_rayleigh_relays}, justified by the reduction of the LOS component due to the directionality \cite{fd:si_model_nlos_matters_bss}. Denoting by $\eta_{si}$ the power gain of the SI channel including the SI cancellation scheme, and by $\kappa_{si}$ its mean value, i.e, $\kappa_{si} = \mathbb{E}[\eta_{si}]$, the power of the remaining SI signal, denoted by $I_{si}$, is given by
\begin{equation}
I_{si} = \eta_{si} P_{ue},
\label{eq:SI}
\end{equation}
where $\eta_{si} \! \sim \! \text{Exp}\left(\frac{1}{\kappa_{si}}\right)$.

\subsection{Resource Allocation and Scheduling}\label{sys:scheduling}
We focus on the downlink of the cellular system, which is isolated from the uplink through frequency division depluxing (FDD), since the uplink performance is not relevant for the considered caching scenario. We further consider an inband overlay scheme for D2D communication \cite{d2d:andrews_overlay_journal}, where a fraction $\chi_{d2d}$ of the overall downlink spectrum $BW$ is reserved for the D2D traffic, justified by the availability of spectrum in the mmWave band. Regarding the scheduling scheme, we consider TDMA scheduling for the active cellular UEs, which is suited to mmWave communication \cite{mmwaves:rappaport_for_tdma}, and uncoordinated D2D comnunication for the D2D UEs, relying on the directionality of the mmWave transmissions for the interference mitigation.

\section{Offloading Analysis}
\label{section:offloading_analysis}
In this section, the DAC and the MPC policies\footnote{Note that the same network topology, as described in Section \ref{sys:network}, has been assumed for both policies to ensure a fair performance comparison.} are compared in terms of their offloading performance, which can be quantified by the \textit{offloading factor} $F$, defined as the ratio of the average offloaded requests (i.e., requests that are not served through cellular connections) to the total content requests in the network, i.e.
\begin{equation}
F \triangleq \frac{\mathbb{E}[\mbox{offloaded requests}]}{\mbox{total requests}}.
\end{equation}
The offloading factor $F$ is derived for each policy as follows:
\begin{itemize}
\item In the MPC policy, a content request can be offloaded only through a cache hit, hence
\begin{equation}
F_{mpc} = h_{mpc}.
\end{equation}

\item In the DAC policy, in addition to a cache hit, a content request of a paired UE can be offloaded through D2D communication, hence
\begin{equation}
F_{dac} = \delta \cdot 2 h_{dac} + (1-\delta) \cdot h_{dac}=(1+\delta) h_{dac}.
\end{equation} 
\end{itemize}

Based on the above, the relative gain of the DAC over the MPC policy in terms of the offloading factor, denoted by $F_{gain}$, is given by
\begin{equation}
F_{gain} = \frac{F_{dac}}{F_{mpc}} = (1+\delta) h_{ratio},
\label{eq:offloading_gain}
\end{equation}
where $h_{ratio}$ represents the ratio of the hit probabilities of the two policies, given by
\begin{equation}
h_{ratio} = \frac{h_{dac}}{h_{mpc}} =  \frac{1}{2}\frac{\sum_{i=1}^{2 K} i^{-\xi}}{\sum_{j=1}^{K} j^{-\xi}}.
\label{hratio}
\end{equation}
We observe that $F_{gain}$ depends on the fraction of the paired UEs $\delta$, the UE cache size $K$ and the content popularity exponent $\xi$, but not the library size $L$. The impact of $K$ and $\xi$ on $h_{ratio}$ and, consequently, $F_{gain}$ is analytically investigated in Proposition 2 that follows.

\begin{proposition}
The ratio of the hit probabilities of the two policies, $h_{ratio}$, decreases monotonically with the popularity exponent $\xi$ and the UE cache size $K$. In addition, the limit of $h_{ratio}$ with high values of $K$ is equal to
\begin{equation}
 \lim_{K \to \infty}  h_{ratio} = \mbox{max}\left(2^{-\xi}, \frac{1}{2}\right).
\label{eq:hitratio_limit}
\end{equation}
\end{proposition}

\begin{IEEEproof}
See Appendix \ref{appendix:hitprob_analysis}.
\end{IEEEproof} 

\begin{figure}[!t]
\centering
\includegraphics[width=3in]{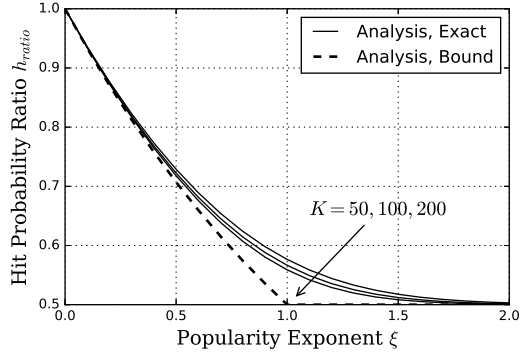}
\caption{The hit probability ratio $h_{ratio}$ in terms of the UE cache size $K$ and the popularity exponent $\xi$.}
\label{fig:hitprob_ratio}
\end{figure}

Proposition 2 implies that $h_{ratio}$ attains its minimum value for $\xi \to \infty$, and its maximum value for $\xi=0$, hence
\begin{equation}
\frac{1}{2}<h_{ratio} \leq 1 \implies \frac{1+\delta}{2}<F_{gain}\leq 1 + \delta.
\end{equation}
This result shows that for $\delta = 1$, representing the case of a fully paired network, the DAC policy always exhibits higher offloading than the MPC policy, while for $\delta = 0$, representing the case of
a fully unpaired network, the converse holds. For an intermediate value of $\delta$, the offloading comparison depends on $\xi$ and $K$ and can be determined through \eqref{eq:offloading_gain}. Finally, in Fig. \ref{fig:hitprob_ratio}, the convergence of $h_{ratio}$ to its limit value for high values of $K$ is depicted. This limit is a lower bound to $h_{ratio}$ and serves as a useful approximation, provided that $\xi$ is not close to 1 because, in this case, the convergence is slow. 

\section{Performance Analysis}
\label{section:performance_analysis}
In this section, the DAC and the MPC policy are characterized in terms of their rate and delay performance. The complementary CDF (CCDF) of the cellular rate is derived in Section \ref{subsection:cellular_rate_arxi}, the CCDF of the D2D rate is derived in Section \ref{subsection:d2d_rate}, and the CDF of the content retrieval delay is derived in Section \ref{subsectio:delay_analysis}.

\subsection{Cellular Rate Analysis}
\label{subsection:cellular_rate_arxi}
Justified by the stationarity of the PPP\cite{stochgeom:haenggi_book}, we focus on a \textit{target UE} inserted at the origin of the network and derive the experienced cellular rate, denoted by $\mathcal{R}_{cell}$, when an uncached content is requested. The rate $\mathcal{R}_{cell}$ is determined by the cellular signal-to-interference-plus-noise ratio (SINR), denoted by $SINR_{cell}$, and the load of the associated cell, denoted by $\mathcal{N}_{cell}$, through the Shannon capacity formula, modified to include the effect of the TDMA scheduling as \cite{ stoch_geom:andrews_cellrate_with_load}
\begin{equation}
\mathcal{R}_{cell}=\frac{BW_{cell}}{\mathcal{N}_{cell}} \log \left(1+SINR_{cell}\right) \mbox{  [bps]}.
\label{eq:cellrate_shannon}
\end{equation}

Based on \eqref{eq:cellrate_shannon}, the distribution of $\mathcal{R}_{cell}$ is derived through the distribution of $SINR_{cell}$ and $\mathcal{N}_{cell}$ as
\begin{align}
&\text{P}(\mathcal{R}_{cell}>\rho)= \text{P}\left(SINR_{cell}>2^{\frac{\rho \mathcal{N}_{cell} }{BW_{cell}}}-1\right) = \nonumber\\ = 
\sum_{n=1}^{\infty} \text{P}&(\mathcal{N}_{cell}=n) \text{P}\left(SINR_{cell}>2^{\frac{\rho n }{BW_{cell}}}-1 \Big| \mathcal{N}_{cell}=n \right) \overset{(i)}\approx \nonumber\\ \approx
\sum_{n=1}^{\infty}  \text{P}&(\mathcal{N}_{cell}=n) \text{P}\left(SINR_{cell}>2^{\frac{\rho n }{BW_{cell}}}-1\right),
\label{def:cellrateccdf}
\end{align}
where (i) follows by treating $SINR_{cell}$ and $\mathcal{N}_{cell}$ as independent random variables\footnote{Please note that $SINR_{cell}$ and $\mathcal{N}_{cell}$ are dependent because the cell load $\mathcal{N}_{cell}$ is correlated with the size of the cell, which in turns influences both the signal received from the associated BS and the interference from the neighboring BSs. Nevertheless, this dependence cannot be modeled analytically, since the relation between the SINR and the cellular size is intractable, and is not expected to have a significant impact on $\mathcal{R}_{cell}$.}. The distributions of $\mathcal{N}_{cell}$ and $SINR_{cell}$ are derived in the following sections.

\subsubsection{Distribution of the cellular load}
\label{subsection:cell_load}
The distribution of $\mathcal{N}_{cell}$ depends on the cell size $A_{cell}$ and the point process of the active cellular UEs, denoted by $\Phi_{cell}$, as follows:
\begin{itemize}
\item Regarding $A_{cell}$, we note that due to the closest BS association scheme, the cells coincide with the Voronoi regions of $\Phi_{bs}$. Although  the area distribution of a typical 2-dimensional Voronoi cell is not known, it can be accurately approximated by \cite{stochgeom:voronoi_area_distribution_approximation}
\begin{equation}
f_{A_{cell}}(a)\approx \frac{(\lambda_{bs} \kappa)^{\kappa} a^{\kappa-1} e^{-\kappa \lambda_{bs} a}}{\Gamma(\kappa)} \,, a>0 \,, \kappa=3.5.
\label{def:typicalvoronoidistr}
\end{equation}
The cell of the target UE, however, is stochastically larger than a randomly chosen cell, since the target UE is more probable to associate with a larger cell, and its area distribution can be derived from \eqref{def:typicalvoronoidistr} as \cite{stochgeom:load_distribution}
\begin{equation}
f_{A_{cell}}(a)=\frac{(\lambda_{bs} \kappa)^{\kappa+1} a^{\kappa} e^{-\kappa \lambda_{bs} a}}{\Gamma(\kappa+1)} \,, a>0 \,, \kappa=3.5.
\label{def:targetvoronoidistr}
\end{equation}

\begin{table}[t]
\renewcommand{\arraystretch}{1.3}
\caption{CELLULAR PROBABILITIES}
\centering
\scalebox{0.7}{
\begin{tabular}{|l | c | c|}
\cline{2-3}
 \multicolumn{1}{c|}{} & $MPC$ & $DAC$\\\hline
$c_u$ & $1-h_{mpc}$ & $1-h_{dac}$ \\\hline
$c_p$ & $1-h_{mpc}$ & $1-2h_{dac}$ \\\hline
\end{tabular}
}
\label{table:cellular_probabilities}
\end{table}

\item Regarding $\Phi_{cell}$, it results from the independent thinning \cite{stochgeom:haenggi_book} of $\Phi_{ue}$, considering the probability of a UE being cellular. This probability is denoted by $c_u$ and $c_p$ for the case of an unpaired and a paired UE respectively, and its values are summarized in Table \ref{table:cellular_probabilities} for the two considered policies. Although $\Phi_{cell}$ is not PPP due to the correlation in the positions of the paired UEs, it can be treated as a PPP with density $\lambda_{cell}$, given by 
\begin{equation}
\lambda_{cell} = \left[  (1-\delta) \cdot c_u + \delta \cdot c_p\right] \lambda_{ue}.
\end{equation}
This approximation is justified by the small cell radius of the mmWave BSs, which is expected to be comparable to the D2D distance of the paired UEs, so that their positions inside the cell are sufficiently randomized.
\end{itemize}

Based on the above, $\mathcal{N}_{cell}$ is approximated with the number of points of one PPP that fall inside the (target) Voronoi cell of another PPP, hence, it is distributed according to the gamma-Poisson mixture distribution \cite{stochgeom:load_distribution}, given by
\begin{equation}
\text{P}(\mathcal{N}_{cell}=n) = \frac{\Gamma(n+\kappa)}{\Gamma(\kappa+1)\Gamma(n)}\mu^{n-1} \left(1-\mu\right)^{\kappa+1} \mbox{, } n \geq 1 \mbox{, }
\label{eq:load_distribution_analysis}
\end{equation}
where
\begin{equation*}
\mu = \frac{\lambda_{cell}}{\kappa \lambda_{bs}+\lambda_{cell}}.
\end{equation*}

\subsubsection{Distribution of the cellular SINR}
\label{subsection:cellular_rate}
The cellular SINR is defined as
\begin{equation}
SINR_{cell} \triangleq \frac{S}{I+N},
\label{eq:cell_sinr}
\end{equation}
where
\begin{itemize}
\item $S$ is a random variable representing the received signal power from the associated BS, which is located at a distance $r$ from the target UE. Assuming that the BS and UE antennas are perfectly aligned, $S$ is given by
\begin{equation}\label{eq:signal_power}
S = \left(\frac{\bar{\lambda}_c}{4\pi}\right)^2 P_{bs}  G_{bs}^{max}G_{ue}^{max} \eta  r_{}^{-a}.
\end{equation}

\item $I$ is a random variable representing the received interference power from the other-cell BSs of $\Phi_{bs}$. Assuming that the UE density is sufficiently high, all the BSs have a UE scheduled and $I$ is given by
\begin{equation}\label{eq:interf_power}
I= \sum_{x \in \Phi_{bs}} \left(\frac{\bar{\lambda}_c}{4\pi}\right)^2 P_{bs} G_x \eta_x  r_x^{-a_x},
\end{equation}
where $r_x$ and $G_x$ are the length and the gain of the interfering link respectively. The latter comprises the antenna gains of the interfering BS and the target UE.

\item $N$ is the noise power at the receiver, given by
\begin{equation}\label{eq:noise_power}
N = N_0 F_N BW_{cell},
\end{equation}
where $N_0$ is the noise power density, $F_N$ is the noise figure of the receiver, and $BW_{cell}$ is the cellular bandwidth.
\end{itemize}

Introducing the normalized quantities
\begin{align}
g_x &\triangleq \frac{G_x}{max(G_x)} = \frac {G_x}{G_{bs}^{max} G_{ue}^{max}}, \nonumber\\
\hat{S} &\triangleq  \eta r_{}^{-a}, \nonumber\\
\hat{I} &\triangleq \sum_{x \in \Phi_{bs}} g_x \eta_x  r_x^{-a_x}, \nonumber\\
\hat{N} &\triangleq \left(\frac{4\pi}{\bar{\lambda}_c}\right)^2 \frac{N_0 F_N BW_{cell}}{P_{bs} G_{bs}^{max} G_{ue}^{max}},
\end{align}
and applying \eqref{eq:signal_power}, \eqref{eq:interf_power}, and \eqref{eq:noise_power} to \eqref{eq:cell_sinr}, the expression for $SINR_{cell}$ is simplified to
\begin{gather}
SINR_{cell} = \frac{\hat{S}}{\hat{I}+\hat{N}} = \nonumber \\
=\frac{\eta  r_{}^{-a}}{\sum_{x \in \Phi_{bs}} g_x \eta_x  r_x^{-a_x}+\left(\frac{4\pi}{\bar{\lambda}_c}\right)^2 \frac{N_0 F_N BW_{cell}}{P_{bs} G_{bs}^{max} G_{ue}^{max}}}.
\label{eq:cell_sinr_normalized}
\end{gather}
The CCDF of $SINR_{cell}$ is subsequently derived as
\begin{align}
& \text{P}\left(SINR_{cell}>T\right) =  \mathbb{E}_{r_{},a,\hat{I}}\left[\text{P}\left(\eta>(\hat{I}+\hat{N})Tr_{}^a\right)\right] \overset{(i)}{=} \nonumber \\
& = \mathbb{E}_{r_{},a,\hat{I}}
 \left[ e^{-(\hat{I}+\hat{N})Tr^a} \right]
\overset{(ii)}{=}\mathbb{E}_{r_{},a} \left[ \mathcal{L}_{\hat{I}}(Tr_{}^a)e^{-\hat{N}Tr_{}^a}\right],
\label{cellular:sinrccdf_full}
\end{align}
where $(i)$ follows from the CCDF of the exponential random variable, and $(ii)$ from the Laplace transform of $\hat{I}$, denoted by $\mathcal{L}_{\hat{I}}(s)$. Considering that the impact of the interference is reduced due to the directionality of the mmWave transmissions, and that the impact of noise is increased due to the high bandwidth of the mmWave band, we assume that the system is \textit{noise-limited}, which means that $SINR_{cell}$ can be approximated by the cellular signal-to-noise ratio (SNR), denoted by $SNR_{cell}$, as
\begin{gather}
\text{P}\left(SINR_{cell}>T\right) \approx
\text{P}\left(SNR_{cell}>T\right) = \mathbb{E}_{r_{},a}\left[e^{-\hat{N}Tr_{}^a}\right] = \nonumber \\
=\int_0^{\infty} \left(e^{-\frac{r}{r_{los}}} e^{-\hat{N}Tr^{a_{L}}}+(1-e^{-\frac{r}{r_{los}}}) e^{-\hat{N}Tr^{a_{N}}} \right) f_r(r)dr,
\label{def:cellular_snr}
\end{gather}
where $f_r(r)$ is given by \eqref{def:rassoc_pdf}. Although the integral in \eqref{def:cellular_snr} cannot be solved in closed form, we present a tight approximation in Proposition 3 that follows.

\begin{proposition}
The CCDF of the cellular SINR can be accurately approximated by
\begin{equation}
\text{P}(SINR_{cell}>T) \approx J_1(T,a_{N})+J_2(T,a_{L})-J_2(T,a_{N}),
\label{eq:cellsinrccdf_final}
\end{equation}
where 
\begin{align}
& J_1(T,a)= \frac{2}{a r_{cell}^2} \left( 
\frac{\gamma\left(\frac{2}{a}, \hat{N} T r_1^{a}\right)}{(\hat{N}T)^{\frac{2}{a}}} -
\frac{\gamma\left(\frac{3}{a}, \hat{N} T r_1^{a}\right)}{r_1 (\hat{N}T)^{\frac{3}{a}}} \right), \\
&J_2(T,a)= \nonumber\\ 
& =\frac{2}{a r_{cell}^2} \left( 
\frac{\gamma\left(\frac{2}{a}, \hat{N} T r_2^{a}\right)}{(\hat{N}T)^{\frac{2}{a}}} -
2\frac{\gamma\left(\frac{3}{a}, \hat{N} T r_2^{a}\right)}{r_2 (\hat{N}T)^{\frac{3}{a}}} +
\frac{\gamma\left(\frac{4}{a}, \hat{N} T r_2^{a}\right)}{r_2^2 (\hat{N}T)^{\frac{4}{a}}}\right),
\end{align}
with
\begin{gather*}
r_1 = \sqrt{3}r_{cell}, \\
r_{2} = \sqrt{6} \sqrt{1-\sqrt{\pi}\frac{r_{cell}}{2 r_{los}} e^{\left(\frac{r_{cell}}{2 r_{los}}\right)^2} \text{erfc}\left(\frac{r_{cell}}{2r_{los}}\right)} r_{cell}.
\end{gather*}
\end{proposition}

\begin{IEEEproof}
See Appendix \ref{appendix:cellular_SINR}.
\end{IEEEproof}

\subsection{D2D Rate Analysis}
\label{subsection:d2d_rate}
Similar to the cellular case, we focus on a paired target UE at the origin and derive the experienced D2D rate, denoted by $\mathcal{R}_{d2d}$, when a content is requested from the D2D peer. The following analysis applies only to the DAC policy, which is distinguished for the HD-DAC and the FD-DAC policy in the following sections.

\subsubsection{Distribution of the D2D rate for the HD-DAC policy}
\label{subsection:hd}
The D2D rate for the HD-DAC policy, denoted by $\mathcal{R}_{d2d}^{hd}$, is determined by the D2D SINR, denoted by $SINR_{d2d}^{hd}$, through the Shannon capacity formula as
\begin{equation}
\mathcal{R}_{d2d}^{hd}=\psi BW_{d2d} \log \left(1+SINR_{d2d}^{hd}\right) \mbox{  [bps]},
\label{eq:hd_d2drate_shannon}
\end{equation}
where $\psi$ denotes the HD factor, equal to $1/2$ when both paired UEs want to transmit. Subsequently, the CCDF of $\mathcal{R}_{d2d}^{hd}$ is determined by the CCDF of $SINR_{d2d}^{hd}$ as
\begin{equation}
\text{P}\left( \mathcal{R}_{d2d}^{hd} > \rho \right) =\text{P}\left(SINR_{d2d}^{hd} >2^{\frac{\rho}{\psi BW_{d2d}}}-1 \right).
\label{eq:hd_d2drate_shannon_ccdf}
\end{equation}
Regarding $SINR_{d2d}^{hd}$, it is defined as:
\begin{equation}
SINR_{d2d}^{hd} \triangleq \frac{S}{I+N},
\label{eq:d2d_hd_sinr}
\end{equation}
where
\begin{itemize}
\item $S$ is a random variable representing the received signal power from the D2D peer, located at a distance $r_{d2d}$ from the target UE. Assuming that the antennas of the two UEs are perfectly aligned, $S$ is given by
\begin{equation}\label{eq:d2d_hd_signal_power}
S = \left(\frac{\bar{\lambda}_c}{4\pi}\right)^2 P_{ue} (G_{ue}^{max})^2 \eta  r_{d2d}^{-a}.
\end{equation}

\item $I$ is a random variable representing the received interference power from all transmitting D2D UEs. Denoting by $\Phi_{d2d}^{hd}$ the point process of the D2D interferers in the HD-DAC policy, $I$ is given by
\begin{equation}\label{eq:d2d_hd_interf_power}
I= \sum_{x \in \Phi_{d2d}^{hd}} \left(\frac{\bar{\lambda}_c}{4\pi}\right)^2 P_{ue} G_x \eta_x  r_x^{-a_x},
\end{equation}
where $r_x$ and $G_x$ are the length and the gain of the interfering link respectively. The latter comprises the antenna gains of the interfering UE and the target UE. Since, in the HD-DAC policy, at most one UE from every D2D pair can transmit, the intensity of $\Phi_{d2d}^{hd}$ is given by 
\begin{equation}
\lambda_{d2d}^{hd}=\left(1-(1-h_{dac})^2 \right) \lambda_{p}=\frac{\delta}{2} h_{dac}\left(2-h_{dac} \right) \lambda_{ue}.
\end{equation}

\item $N$ is the noise power at the receiver, which depends on the D2D bandwidth $BW_{d2d}$ and is given by
\begin{equation}\label{eq:d2d_hd_noise_power}
N = N_0 F_N BW_{d2d}.
\end{equation}
\end{itemize}

Introducing the normalized quantities
\begin{align}
g_x & \triangleq \frac{G_x}{\mbox{max}(G_x)} = \frac {G_x}{(G_{ue}^{max})^2}, \nonumber\\
\hat{S} & \triangleq  \eta r_{d2d}^{-a}, \nonumber\\
\hat{I} & \triangleq \sum_{x \in \Phi_{bs}} g_x \eta_x  r_x^{-a_x}, \nonumber\\
\hat{N} & \triangleq \left(\frac{4\pi}{\bar{\lambda}_c}\right)^2 \frac{N_0 F_N BW_{d2d}}{P_{ue} (G_{ue}^{max})^2},
\label{eq:hd_d2d_normalizations}
\end{align}
and applying \eqref{eq:d2d_hd_signal_power}, \eqref{eq:d2d_hd_interf_power}, and \eqref{eq:d2d_hd_noise_power} to \eqref{eq:d2d_hd_sinr}, the expression for $SINR_{d2d}^{hd}$ is simplified to
\begin{equation}
SINR_{d2d}^{hd} = \frac{\hat{S}}{\hat{I}+\hat{N}} = \frac{\eta  r_{d2d}^{-a}}{\sum_{x \in \Phi_{d2d}^{hd}} g_x \eta_x  r_x^{-a_x}+\left(\frac{4\pi}{\bar{\lambda}_c}\right)^2 \frac{N_0 F_N BW_{d2d}}{P_{ue}(G_{ue}^{max})^2}}.
\label{eq:d2d_hd_sinr_normalized}
\end{equation}
The  CCDF of $SINR_{d2d}^{hd}$ is derived similarly to \eqref{cellular:sinrccdf_full} as
\begin{align}
&\text{P}(SINR_{d2d}^{hd}>T) = \mathbb{E}_{r_{d2d},a} \left[ 
\mathcal{L}_{\hat{I}}^{hd}(s) 
e^{-\hat{N}s}\right], \,s=T r_{d2d}^a,
\label{eq:d2d_hd_sinrccdf}
\end{align}
where $\mathcal{L}_{\hat{I}}^{hd}(s)$ is the Laplace transform of the interference in the HD-DAC policy, and the expectation over $a$ and $r_{d2d}$ is computed through \eqref{eq:plos} and \eqref{eq:d2d_displacement} respectively. In contrast to the cellular case, the contribution of the interference in $SINR_{d2d}^{hd}$ is not negligible, even with directionality, due to the lower bandwidth expected for D2D communication, thus, $\mathcal{L}_{\hat{I}}^{hd}(s)$ is evaluated according to Proposition 4 that follows.

\begin{proposition}
The Laplace transform of the D2D interference in the HD-DAC policy, $\mathcal{L}_{\hat{I}}^{hd}(s)$, is given by
\begin{align}
\mathcal{L}_{\hat{I}}^{hd}(s)
\approx
 e^{-\pi \delta h_{dac}(2-h_{dac})\lambda_{ue} \mathbb{E}_g
 \left[
 J_3\left(s,a_N \right) + J_4\left(s,a_N ;k\right) - J_4\left(s,a_L;k\right)
 \right]},
\label{eq:d2d_hd_interf_laplace}
\end{align}
where
\begin{gather}
J_3(s, a) = \frac{1}{2} \Gamma\left(1-\frac{2}{a}\right)\Gamma\left(1+\frac{2}{a}\right) g^{\frac{2}{a}} s^{\frac{2}{a}}, \nonumber\\
J_4(s, a; k) = \sum_{l=0}^k \binom{k}{l} (-1)^l 
\frac{r_4^{a+2}  {}_2 F_1\left(1,1+\frac{l+2}{a};2+\frac{l+2}{a};-\frac{r_4^a}{gs}\right)}{(l+a+2)gs},  \nonumber\\
r_4 = \sqrt{(k+1)(k+2)} r_{los},
\label{eq:d2d_hd_interf_laplace_terms}
\end{gather}
$k$ denotes the order of the approximation, and the averaging is taken over the discrete random variable $g$ with distribution
\begin{subnumcases}{g=}
1 & 
with prob $\frac{\Delta\theta_{ue}^2}{4\pi^2}$ \\
\frac{G_{ue}^{min}}{G_{ue}^{max}} & 
with prob $2 \frac{\Delta\theta_{ue}(2\pi-\Delta\theta_{ue})}{4\pi^2}$ \\
\left(\frac{G_{ue}^{min}}{G_{ue}^{max}}\right)^2 & with prob $\frac{(2\pi-\Delta\theta_{ue})^2}{4\pi^2}$
\end{subnumcases}
\end{proposition}

\begin{IEEEproof}
See Appendix \ref{appendix:hd-dac}.
\end{IEEEproof}

As $k \to \infty$, more terms are added in the summation and the approximation becomes exact. Combining \eqref{eq:d2d_hd_interf_laplace} with \eqref{eq:d2d_hd_sinrccdf} into \eqref{eq:hd_d2drate_shannon_ccdf} yields the CCDF of $\mathcal{R}_{d2d}^{hd}$ where the final integration over $r_{d2d}$ can be evaluated numerically.

\subsubsection{Distribution of the D2D rate for the FD-DAC policy}
\label{subsection:fd}
As in the case of the HD-DAC policy, the D2D rate for the FD-DAC policy, denoted by $\mathcal{R}_{d2d}^{fd}$, is determined by the D2D SINR, denoted by $SINR_{d2d}^{fd}$, through the Shannon capacity formula as
\begin{equation}
\mathcal{R}_{d2d}^{fd}=BW_{d2d} \log \left(1+SINR_{d2d}^{fd}\right) \mbox{  [bps]}.
\label{eq:fd_d2drate_shannon}
\end{equation}
Subsequently, the CCDF of $\mathcal{R}_{d2d}^{fd}$ is derived from the CCDF of $SINR_{d2d}^{fd}$ as
\begin{equation}
\text{P}\left( \mathcal{R}_{d2d}^{fd} > \rho \right) =\text{P}\left(SINR_{d2d}^{fd} >2^{\frac{\rho}{BW_{d2d}}}-1 \right).
\label{eq:fd_d2d_rate_shannon_ccdf}
\end{equation}
Regarding $SINR_{d2d}^{fd}$, it is defined as
\begin{equation}
SINR_{d2d}^{fd} \triangleq \frac{S}{I+I_{si}+N},
\label{eq:d2d_fd_sinr}
\end{equation}
where
\begin{itemize}
\item $S$ is a random variable representing the received signal power from the D2D peer, given by \eqref{eq:d2d_hd_signal_power}.
\item $I_{si}$ is a random variable representing the SI power when the target UE operates in FD mode, given by \eqref{eq:SI}.
\item $I$ is a random variable representing the received interference power from all transmitting D2D UEs, given by  
\begin{align}
I = &\sum_{x_1 \in \Phi_{p}^{(1)}} \left(\frac{\bar{\lambda}_c}{4\pi}\right)^2 P_{ue} \psi_{x_1} G_{x_1} \eta_{x_1}  r_{x_1}^{-a_{x_1}}+ \nonumber\\
+ & \sum_{x_2 \in \Phi_{p}^{(2)}} \left(\frac{\bar{\lambda}_c}{4\pi}\right)^2 P_{ue} \psi_{x_2} G_{x_2} \eta_{x_2}  r_{x_2}^{-a_{x_2}},
\end{align}
where $\Phi_{p}^{(1)}$ and $\Phi_{p}^{(2)}$ are the point processes of the paired UEs, and $\psi_{x}$ denotes the indicator variable for the event that the UE at position $x$ transmits.
\item $N$ is the noise power at the receiver, given by \eqref{eq:d2d_hd_noise_power}.
\end{itemize}

Defining $g$, $\hat{S}$ and $\hat{N}$ as in \eqref{eq:hd_d2d_normalizations} and introducing
\begin{align}
\hat{I} &= \sum_{x \in \Phi_{p}^{(1)}} \psi_x g_x \eta_x  r_x^{-a_x}+\sum_{y \in \Phi_{p}^{(2)}} \psi_y g_y \eta_y  r_y^{-a_y}, \nonumber\\
\hat{I_{si}} &= \left(\frac{4\pi}{\bar{\lambda}_c}\right)^2 \frac{\eta_{si}}{(G_{ue}^{max})^2}, 
\end{align}
the CCDF of $SINR_{d2d}^{fd}$ is derived similarly to \eqref{eq:d2d_hd_sinrccdf} as
\begin{equation}
\text{P}\left(SINR_{d2d}^{fd}>T\right) = \mathbb{E}_{r_{d2d},a}
\left[ \mathcal{L}_{\hat{I}}^{fd}(s)  \mathcal{L}_{\hat{I_{si}}}(s)
 e^{-\hat{N}s}\right],\, s=T r_{d2d}^a,
 \label{d2d:sinrccdf_fd_full}
\end{equation}
where $\mathcal{L}_{\hat{I}}^{fd}(s)$ and $\mathcal{L}_{\hat{I_{si}}}(s)$ are the Laplace transforms of the external D2D interference and the SI respectively. Recalling that $\eta_{si} \! \sim \! \text{Exp}\left(\frac{1}{\kappa_{si}}\right)$, $\mathcal{L}_{\hat{I_{si}}}(s)$ is derived through the Laplace transform of the exponential random variable as
\begin{equation}
\mathcal{L}_{\hat{I_{si}}}(s)=
\mathbb{E} \left[
e^{- \left( \frac{4\pi}{ \bar{\lambda}_c G_{ue}^{max}}\right)^2 \eta_{si} s}
\right]
= \frac{1}{1+\left( \frac{4\pi}{ \bar{\lambda}_c G_{ue}^{max}}\right)^2 \frac{s}{\kappa_{si}}},
\label{eq:d2d_fd_laplace_si}
\end{equation}
while $\mathcal{L}_{\hat{I}}^{fd}(s)$ is derived in Proposition 5 that follows.

\begin{proposition}
The Laplace transform of the D2D interference in the FD-DAC policy, $\mathcal{L}_{\hat{I}}^{fd}(s)$, can be bounded as
\begin{align}
&\mathcal{L}_{\hat{I}}^{fd}(s) \geq e^{-\pi \delta \lambda_{ue} h_{dac} 2  \mathbb{E}_g
 \left[ J_3\left(s,a_N \right) + J_4\left(s,a_N ;k\right) - J_4\left(s,a_L;k\right)\right]}, \label{eq:d2d_fd_interf_laplace_lb}
\\
& \mathcal{L}_{\hat{I}}^{fd}(s) \leq
e^{-\pi \delta \lambda_{ue} h_{dac}  \mathbb{E}_g
 \left[ J_3\left(2s,a_N \right) + J_4\left(2s,a_N ;k\right) - J_4\left(2s,a_L;k\right)\right]},
\label{eq:d2d_fd_interf_laplace_ub}
\end{align}
where $J_3\left(s,a \right)$ and $J_4\left(s,a ;k\right)$ are given by \eqref{eq:d2d_hd_interf_laplace_terms}.
\end{proposition}

\begin{IEEEproof}
See Appendix \ref{appendix:fd-dac}.
\end{IEEEproof}

Combining \eqref{eq:d2d_fd_interf_laplace_lb} and \eqref{eq:d2d_fd_interf_laplace_ub} with \eqref{eq:d2d_fd_laplace_si} into \eqref{d2d:sinrccdf_fd_full}, and applying the result to \eqref{eq:fd_d2d_rate_shannon_ccdf}, yields two bounds for the CCDF of $\mathcal{R}_{d2d}^{fd}$. 

\subsection{Delay Analysis}
\label{subsectio:delay_analysis}
In this section, we characterize the delay performance of the MPC and the DAC policies through the \textit{content retrieval delay}, denoted by $D$ and defined as the delay experienced by a UE when retrieving a requested content from any available source. In the case of a cache hit, $D$ is zero, while in the cellular and the D2D case, it coincides with the transmission delay of the content to the UE\footnote{Additional delays caused by the retrieval of the content through the core network are beyond the scope of this work.}. The CDFs of $D$ for the MPC and the DAC policy are derived as follows:
\begin{itemize}

\item For the MPC policy, the requested content is retrieved from the local cache with probability $h_{mpc}$, or from the BS with probability $1-h_{mpc}$, hence 
\begin{equation}
\text{P} \left(D<d \right)=h_{mpc}+(1-h_{mpc})\text{P} \left(\mathcal{R}_{cell} >\frac{\sigma_{file}}{d} \right),
\label{delay:mpc}
\end{equation}
where the CCDF of $\mathcal{R}_{cell}$ is given by \eqref{def:cellrateccdf}.

\item For the DAC policy, the case of the paired and the unpaired UE must be distinguished, since the unpaired UE lacks the option for D2D communication. For a paired UE, the requested content is retrieved from the local cache with probability $h_{dac}$, from the D2D peer with probability $h_{dac}$, or from the BS with probability $1-2h_{dac}$, while, for an unpaired UE, the requested content is retrieved from the local cache with probability $h_{dac}$, or from the BS with probability $1-h_{dac}$, yielding
\begin{align}
\text{P} &\left(D<d \right) = h_{dac}+ \delta h_{dac} \text{P} \left(\mathcal{R}_{d2d} >\frac{\sigma_{file}}{d} \right)+  \nonumber\\  &
+(1-h_{dac}-\delta h_{dac})  \text{P} \left(\mathcal{R}_{cell} >\frac{\sigma_{file}}{d} \right),
\label{delay:dac}
\end{align}
where the CCDF of $\mathcal{R}_{cell}$ is given by \eqref{def:cellrateccdf}, and the CCDF of $\mathcal{R}_{d2d}$ is given by \eqref{eq:hd_d2drate_shannon_ccdf} for the HD-DAC policy and by  \eqref{eq:fd_d2d_rate_shannon_ccdf} for the FD-DAC policy. 
\end{itemize}

\section{Results}
\label{section:results}
In this section, we compare the DAC and the MPC policies in terms of the offloading factor and the 90-th percentile of the content retrieval delay analytically and through Monte-Carlo simulations. Towards this goal, we present the simulation parameters in Section \ref{section:network_setup}, the results for the offloading in Section \ref{section:offloading_results}, and the results for the content retrieval delay in Section \ref{section:delay_results}.

\subsection{Simulation Setup}
\label{section:network_setup}
\begin{table}[t]
\renewcommand{\arraystretch}{1.3}
\caption{SIMULATION PARAMETERS}
\centering
\scalebox{0.7}{
\begin{tabular}{|l | c || l | c|}
\hline
$\lambda_{bs}$ & $127  \mbox{ BSs/km}^2$ &   $N_0$  &  -174 dBm/Hz \\
$\lambda_{ue}$ & $1270  \mbox{ UEs/km}^2$ & $F_N$  & 10 dB \\
$\delta$  & 0.5, 0.75, 1   & $\Delta\theta_{ue}$ & 30$^o$ \\
$r_{d2d}^{max}$  & 15 m    & $\Delta\theta_{bs}$ & 10$^o$\\\cline{0-1}
$f_{c}$  &  28 GHz         & $G_{bs}^{max}, $ & 18 dB\\
$BW$  & 2 GHz              & $G_{bs}^{min}$  & -2 dB \\
$\chi_{d2d}$  & 20\%       & $G_{ue}^{max}$ & 9 dB \\
$r_{los}$ & 30 m           & $G_{ue}^{min}$ & -9 dB\\\cline{3-4}
$a_{L}$  & 2             & $\sigma_{file}$ & 100 MBs \\
$a_{N}$  & 3            & $L$ & 1000\\
$P_{bs}$ & 30 dBm          & $K$ & 50, 100, 200\\
$P_{ue}$ & 23 dBm          & $\xi$ & variable \\
$ \kappa_{si}$ & -80 dB     &  & \\ \hline
\end{tabular}
}
\label{table:parameters} 
\end{table}

For the simulation setup of the DAC and the MPC policy, we consider a mmWave system operating at the carrier frequency $f_c$ of 28 GHz, which is chosen due to its favorable propagation characteristics \cite{general:rappaport_mmwaves_itwillwork} and its approval for 5G deployment by the FCC \cite{general:mmwaves_fcc_report}. Regarding the network topology, we consider a high BS density $\lambda_{bs}$ corresponding to an average cell radius $r_{cell}$ of 50 m, which is consistent with the trends in the densification of future cellular networks and the average LOS radius $r_{los}$ of the mmWave frequencies in urban environments \cite{mmwaves:andrews_kulkarni_rate_trends_for_blockage_param_values}. The latter is chosen to be 30 m, based on the layout for the Chicago and the Manhattan area \cite{mmwaves:andrews_kulkarni_rate_trends_for_blockage_param_values}. Regarding the antenna model of the BSs and the UEs, the gains and the beamwidths are chosen according to typical values of the literature \cite{mmwaves:andrews_tractable_selfbackhauled, d2dmmwaves:heath_bai_adhoc_older}, considering lower directionality for the UEs due to the smaller number of antennas that can be installed in the UE devices\footnote{A planar phased array with a beamwidth of $30^o$ can be constructed with 12 antenna elements\cite{general:skolnik}, requiring an area of approximately 3 cm$^2$ at 28 GHz, which is feasible in modern UE devices.}. Regarding the caching model, we consider a library of 1000 files of size 100 MBs and three cases for the UE cache size: i) $K=50$, ii) $K=100$, and iii) $K=200$, corresponding to the 5\%, 10\%, and 20\% percentages of the library size respectively. The simulation parameters are summarized in Table \ref{table:parameters}.

\subsection{Offloading Comparison}
\label{section:offloading_results}
As shown analytically in Section \ref{section:offloading_analysis}, the offloading gain of the DAC policy over the MPC policy $F_{gain}$ increases monotonically with the UE pairing probability $\delta$, and decreases monotonically with the UE cache size $K$ and the content popularity $\xi$, while it is not affected by the library size $L$. In this section, we validate the impact of $\delta$, $K$, and $\xi$ on $F_{gain}$ by means of simulations. 

\begin{figure}[!t]
\centering
\includegraphics[width=3in]{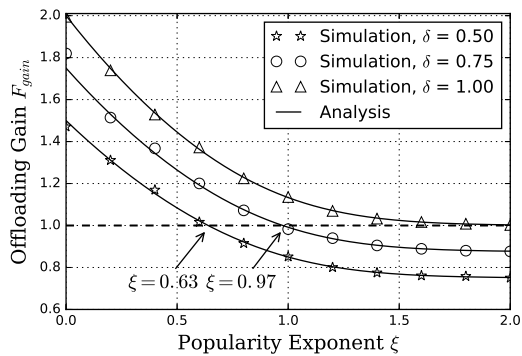}
\caption{The offloading gain of the DAC policy over the MPC policy, $F_{gain}$, in terms of the content popularity exponent $\xi$.}
\label{fig:offloading_gain_versus_xi}
\end{figure}

In Fig. \ref{fig:offloading_gain_versus_xi}, we plot $F_{gain}$ in terms of $\xi$ for $K=100$ and for $\delta= 0.5,\mbox{ } 0.75,\mbox{ } 1$, corresponding to three different percentages of paired UEs inside the network. We observe that the simulation results validate the monotonic increase and decrease of $F_{gain}$ with $\delta$ and $\xi$ respectively. The former is attributed to the higher availability of D2D pairs, which improves the opportunities for offloading in the DAC policy and does not affect the MPC policy, while the latter is attributed to the increasing gap in the hit probabilities of the two policies, as illustrated with the decrease of $h_{ratio}$ with $\xi$ in Fig. \ref{fig:hitprob_ratio} of Section \ref{section:offloading_analysis}. Based on the above, we observe that the maximum offloading gain of the DAC over the MPC policy is equal to 2 and it is achieved when $\delta=1$ and $\xi=0$, which corresponds to the case of a fully paired network and uniform content popularity respectively. For $\delta=1$, we further observe that the DAC policy outperforms the MPC policy regardless of the value of $\xi$, while for lower values of $\delta$, the DAC policy is superior only when $\xi<0.63$ for $\delta=0.75$, and when $\xi<0.97$ for $\delta=0.5$. Based on these observations, we can generalize that for a network with $\delta < 1$ the DAC policy offers higher offloading than the MPC policy for $\xi$ up to a threshold value, which decreases with $\delta$.

\begin{figure}[!t]
\centering
\includegraphics[width=3in]{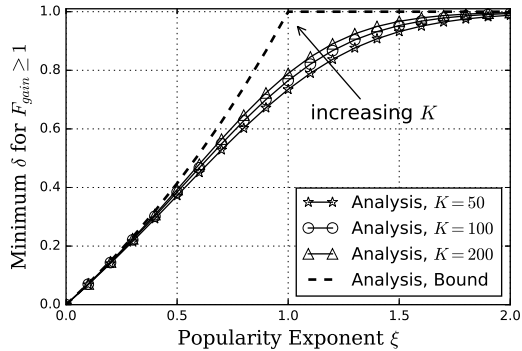}
\caption{The minimum fraction of pairs ($\delta$) required for the DAC policy to achieve higher offloading than the MPC policy in terms of the content popularity exponent $\xi$.}
\label{fig:offloading_min_delta}
\end{figure}

In Fig. \ref{fig:offloading_min_delta}, we plot the minimum $\delta$ that is required for the DAC policy to outperform the MPC policy, in terms of $\xi$ and for $K=50,\mbox{ }100,\mbox{ }200$. We can observe that the requirements for $\delta$ become more stringent with increasing $\xi$ and $K$, which widen the gap between the hit probability of the two policies, but the impact of $K$ is weaker than the impact of $\xi$, which is attributed to the low sensitivity of $h_{ratio}$ with $K$. This behavior can be explained with the bound of $h_{ratio}$ in \eqref{eq:hitratio_limit}, which represents the limit of $h_{ratio}$ when $K \to \infty$. The minimum $\delta$ for $K \to \infty$ is also depicted in Fig. \ref{fig:offloading_min_delta}, as well as the convergence of the other curves to it. When $\xi<0.5$ or $\xi>1.5$, the gap between the curves for finite $K$ are close to the bound, because $h_{ratio}$ converges quickly to its limit value. In contrast, when $0.5<\xi<1.5$, the gap between the curves and the bound is wider, because $h_{ratio}$ converges slowly to its limit value. Due to the slow convergence, for practical values of $K$, similar to ones considered in this work, $h_{ratio}$ is insensitive to $K$.

\begin{figure}[!t]
\centering
\includegraphics[width=3in]{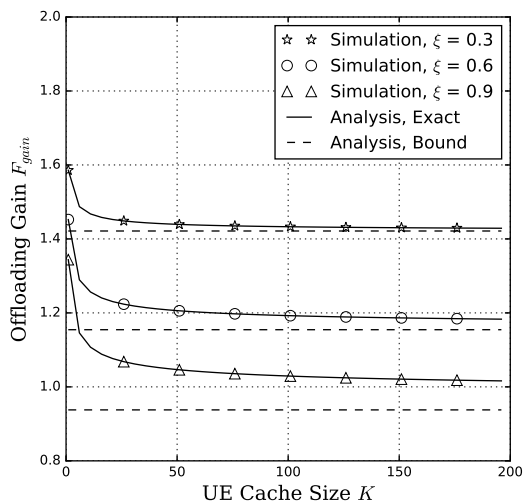}
\caption{The offloading gain of the DAC policy over the MPC policy, $F_{gain}$, in terms of the UE cache size $K$.}
\label{fig:offloading_gain_versus_K}
\end{figure}

In Fig. \ref{fig:offloading_gain_versus_K}, we plot $F_{gain}$ in terms of $K$ for $\delta=0.75$ and $\xi = 0.3, \mbox{ } 0.6, \mbox{ }0.9$. We observe that, as $K$ increases, $F_{gain}$ decreases fast at low values of $K$ and, afterwards, tends slowly to its limit value, calculated by applying \eqref{eq:hitratio_limit} to \eqref{eq:offloading_gain}. For $\xi = 0.9$, the gap between the curve and the limit is high because of the slow convergence of \eqref{eq:hitratio_limit}, validating that $F_{gain}$ is insensitive to $K$, provided that $K$ is sufficiently high. In contrast, lower values of $K$ favor the DAC over the MPC policy.

\subsection{Delay Comparison}
\label{section:delay_results}
In this section, we validate the analytical expressions of Section \ref{section:performance_analysis} and compare the two caching policies in terms of the 90-th percentile of the content retrieval delay.

\subsubsection{Performance of the HD-DAC policy}
\label{sec:performance_HD_DAC}

\begin{figure}[!t]
\centering
\subfloat[Rate]{
\includegraphics[width=1.6in]{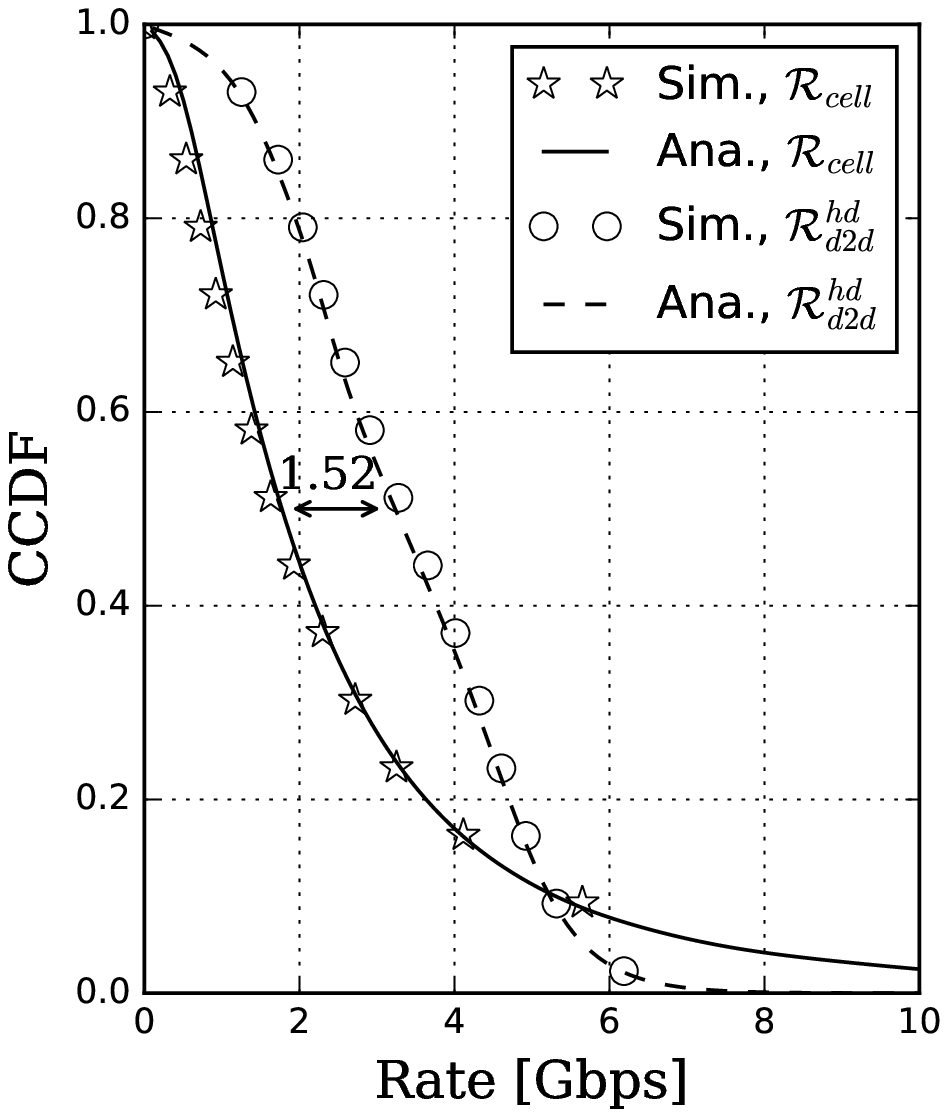}
\label{fig:hd_dac_rate}
} \hfil
\subfloat[Delay]{
\includegraphics[width=1.6in]{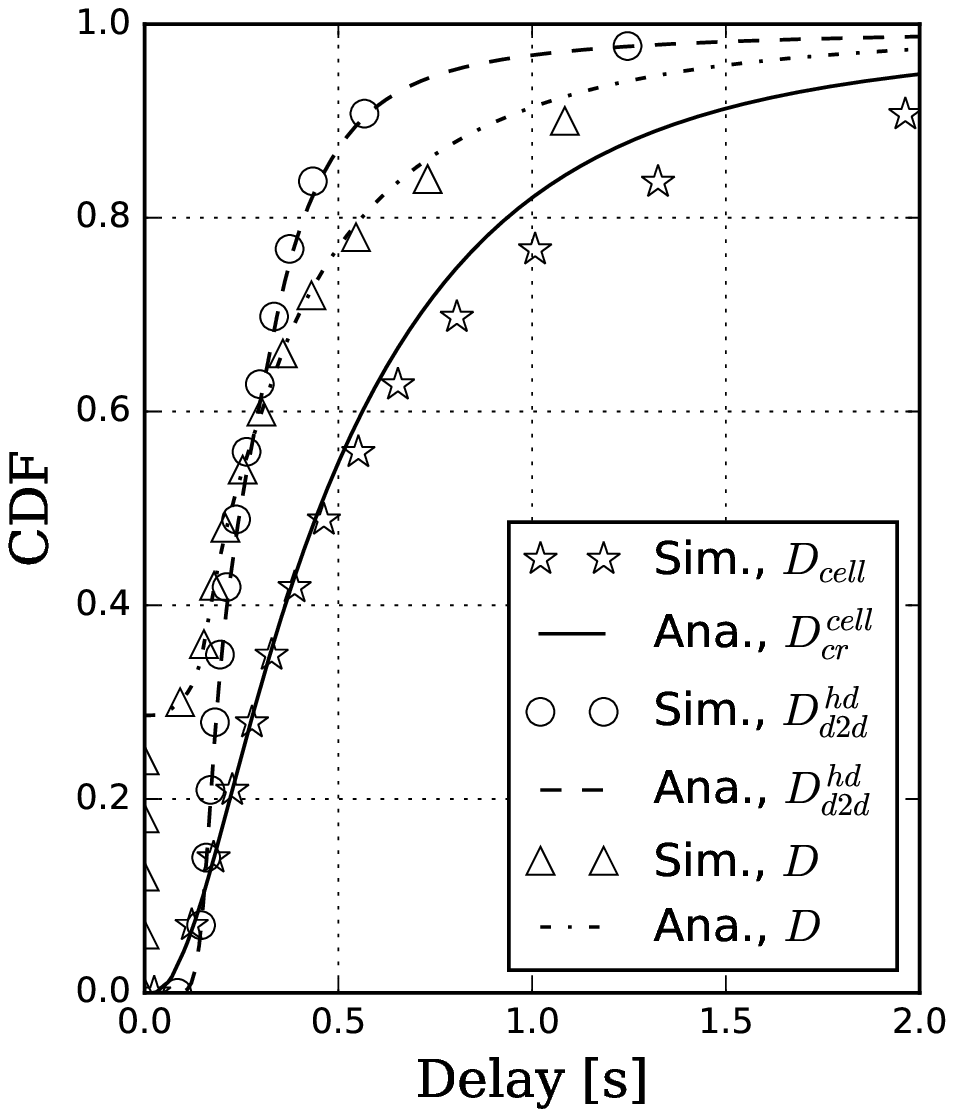}
\label{fig:hd_dac_delay}
}
\caption{Rate and delay performance of the HD-DAC policy for $\delta=1$, $K=200$ and $\xi=0.4$ (Ana. stands for Analysis and Sim. for Simulation).}
\label{fig:hd_dac}
\end{figure}

In Fig. \ref{fig:hd_dac_rate}, we illustrate for the HD-DAC policy the CCDFs of the cellular rate $\mathcal{R}_{cell}$ and the D2D rate $\mathcal{R}_{d2d}^{hd}$, derived through analysis and simulations, for $\delta=1$, $K=200$, and $\xi=0.4$. A second order approximation ($k=2$) was sufficient for $\mathcal{L}_{\hat{I}}^{hd}(s)$ in \eqref{eq:d2d_hd_interf_laplace_terms}. We observe that $\mathcal{R}_{d2d}^{hd}$ is stochastically larger than $\mathcal{R}_{cell}$ for rates below 5 Gbps, yielding an improvement of 1.52 Gbps in the 50-th percentile, which means that the D2D UEs experience a rate that is higher than the cellular rate by at least 1.5 Gbps for the 50\% of the time. This improvement creates strong incentives for the UEs to cooperate and is attributed to the small D2D distance between the D2D UEs and the reduction of $\mathcal{R}_{cell}$ due to the TDMA scheduling. In contrast, the cellular UEs are more probable to experience rates above 5 Gbps, owing to the high difference between the cellular and the D2D bandwidth. Specifically, it is possible for a cellular UE to associate with a BS with low or even zero load and fully exploit the high cellular bandwidth, while a D2D UE is always limited by the $20\%$ fraction of bandwidth that is reserved for D2D communication.

In Fig. \ref{fig:hd_dac_delay}, we illustrate for the HD-DAC policy the CDFs of the cellular delay $D_{cell}$, the D2D delay $D_{d2d}^{hd}$, and the total delay $D$ that is experienced by a UE without conditioning on its content request. We observe that $D_{d2d}^{hd}$ is significantly lower than $D_{cell}$, which is consistent with Fig. \ref{fig:hd_dac_rate}, while the curve of $D$ is initiated at the value $0.286$ due to the zero delay of cache-hits. We further observe that the simulations for $D_{cell}$ do not match the theoretical curve as tightly as in the case of $\mathcal{R}_{cell}$, which is attributed to the reciprocal relation between the rate and the delay that magnifies the approximation error for the delay. Nevertheless, the match is improved in the case of the total delay due to the contribution of the D2D delay, which is approximated more accurately.

\subsubsection{Performance of the FD-DAC policy}
\label{sec:performance_FD_DAC}
\begin{figure}[!t]
\centering
\subfloat[Rate]{
\includegraphics[width=1.6in]{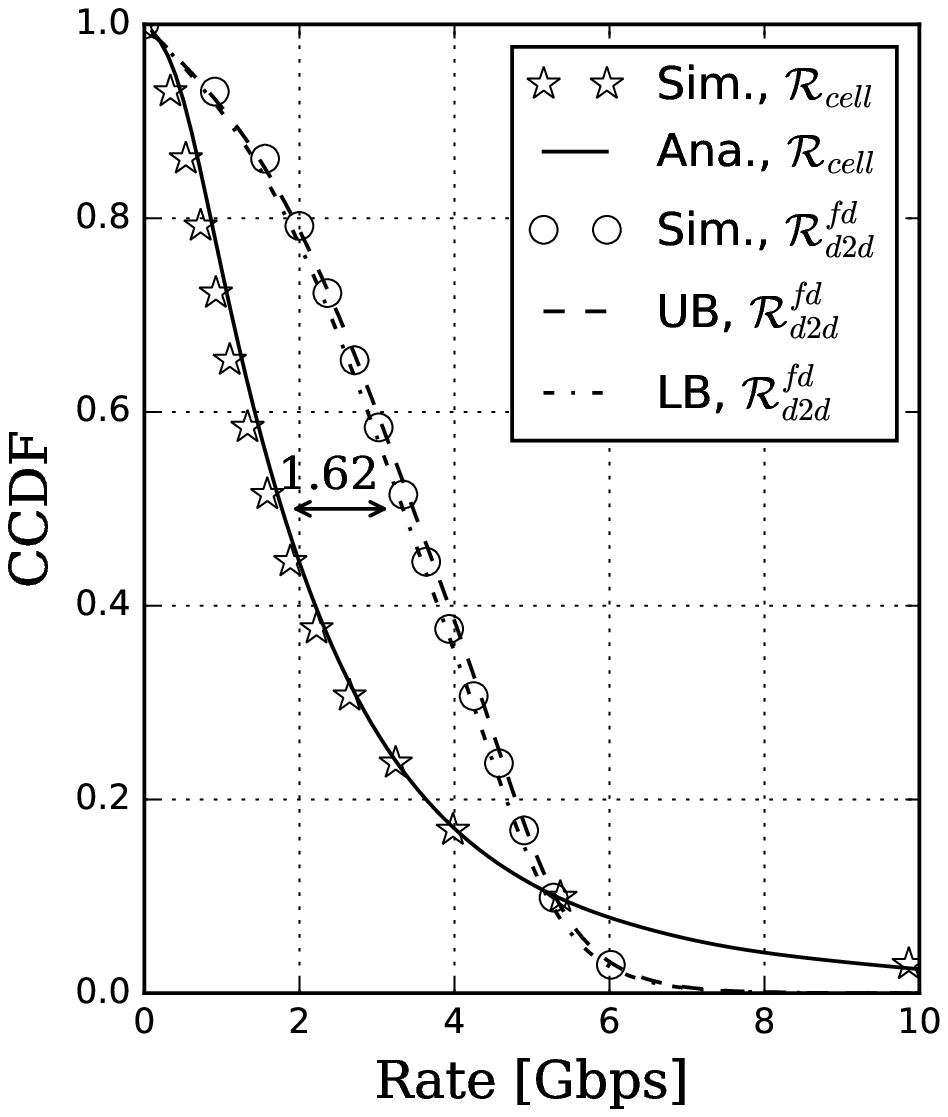}
\label{fig:fd_dac_rate}
} \hfil
\subfloat[Delay]{
\includegraphics[width=1.6in]{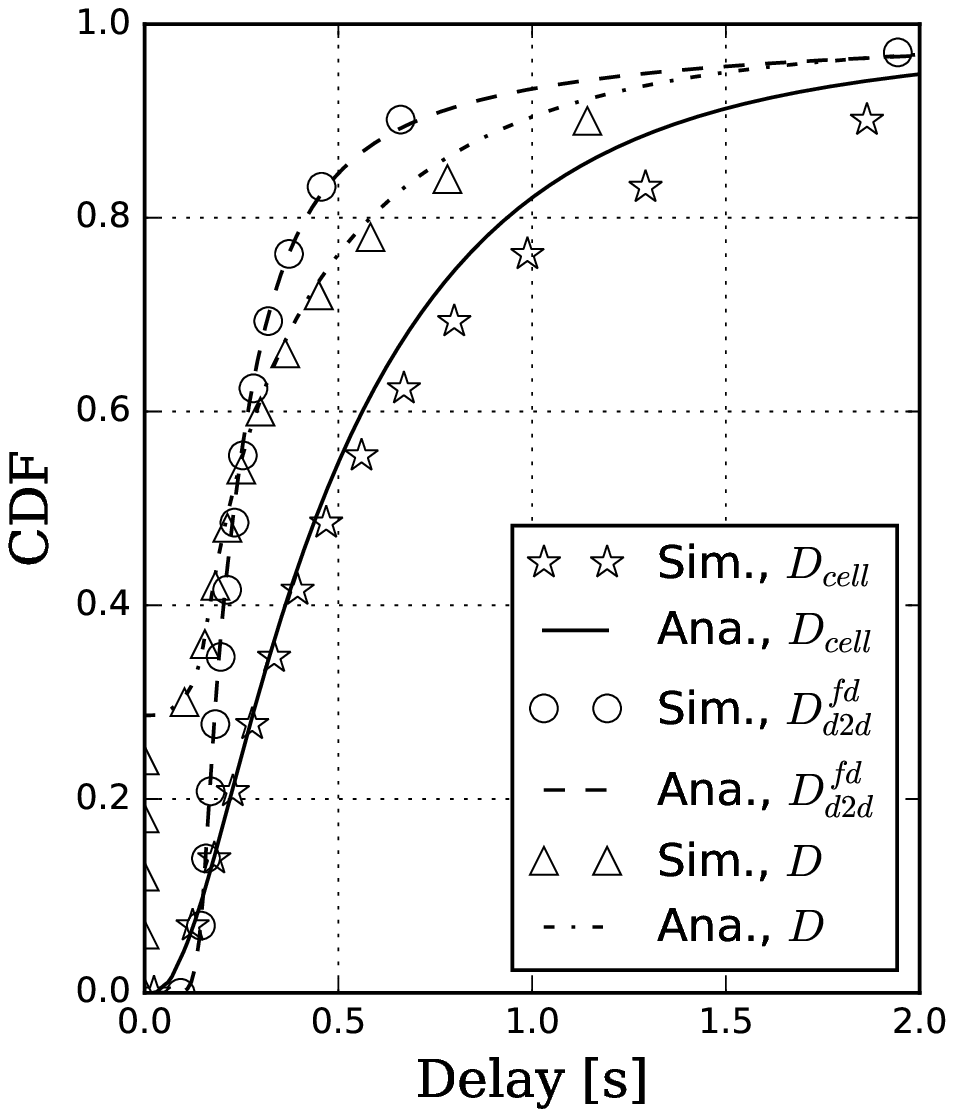}
\label{fig:fd_dac_delay}
}
\caption{Rate and delay performance of the FD-DAC policy for $\delta=1$, $K=200$, and $\xi=0.4$ (Ana. stands for Analysis, Sim. for Simulation, UB for Upper Bound, and LB for Lower Bound).}
\label{fig:fd_dac}
\end{figure}

In Fig. \ref{fig:fd_dac}, we illustrate for the FD-DAC policy the rate and the delay distribution for $\delta=1$, $K=200$, $\xi=0.4$, and a second order approximation for $\mathcal{L}_{\hat{I}}^{fd}(s)$. As seen in Fig. \ref{fig:fd_dac_rate}, both bounds for the CCDF of $\mathcal{R}_{d2d}^{fd}$ are very close to the simulation curve, hence, only the upper bound is considered for $D_{d2d}^{fd}$ in Fig. \ref{fig:fd_dac_delay}.
Compared with the HD-DAC policy, the FD-DAC policy yields a minor improvement in the 50-th percentile of $\mathcal{R}_{d2d}^{fd}$, which is higher than the percentile of $\mathcal{R}_{cell}$ by 1.62 Gbps, that is attributed to the absence of the HD factor that decreases $\mathcal{R}_{d2d}^{hd}$ by half. Nevertheless, the probability of bidirectional content exchange, equal to $0.08$ for the considered parameters, is small to significantly influence the results. The same observation holds for the CDFs of the content retrieval delay.

\begin{figure}[!t]
\centering
\subfloat[Rate]{
\includegraphics[width=1.6in]{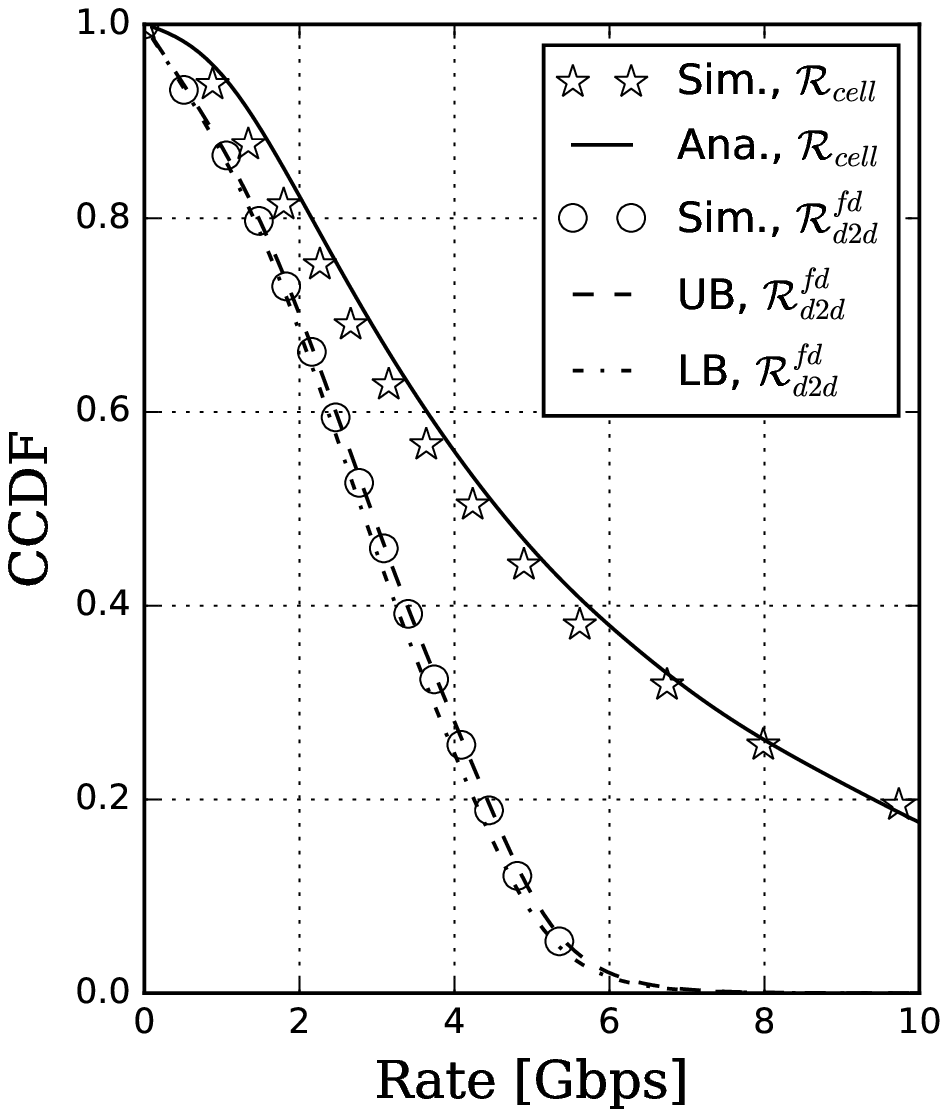}
\label{fig:fd_dac_rate_high_xi}
} \hfil
\subfloat[Delay]{
\includegraphics[width=1.6in]{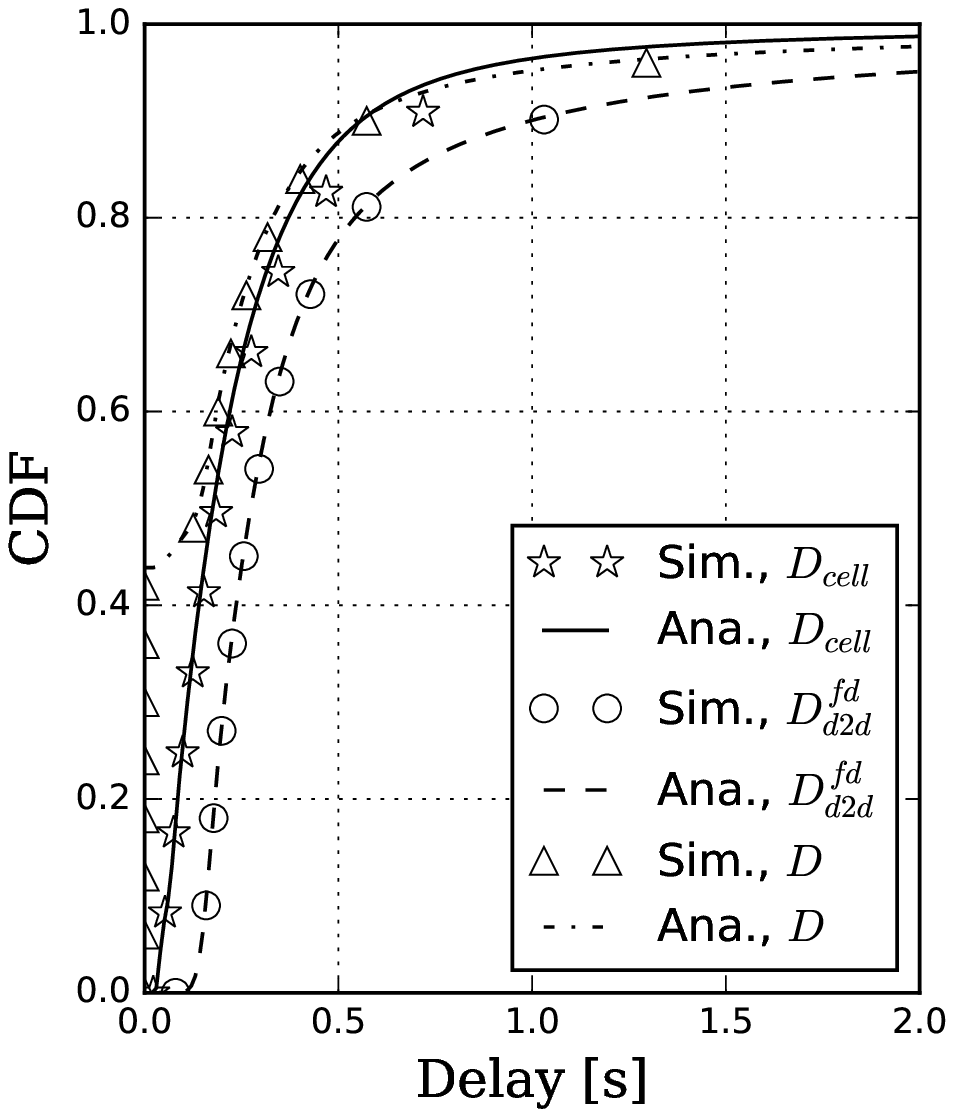}
\label{fig:fd_dac_delay_high_xi}
}
\caption{Rate and delay performance of the FD-DAC policy for $\delta=1$, $K=200$, and $\xi=1.0$ (Ana. stands for Analysis, Sim. for Simulation, UB for Upper Bound, and LB for Lower Bound).}
\label{fig:fd_dac_high_xi}
\end{figure}

Motivated by the previous observation, in Fig. \ref{fig:fd_dac_high_xi}, we illustrate for the FD-DAC policy the rate and the content retrieval delay for $\xi=1.0$, in which case $h_{dac}=0.44$, resulting in a non-negligible probability for bidirectional content exchange. As seen in Fig. \ref{fig:fd_dac_rate_high_xi}, $\mathcal{R}_{d2d}^{fd}$ is reduced due to the higher D2D interference, while $\mathcal{R}_{cell}$ is significantly improved due to the higher offloading. Consequently, $\mathcal{R}_{d2d}^{fd}$ is higher than $\mathcal{R}_{cell}$, and the total delay is determined by the cache hits and the curve of the cellular delay, 
as seen in Fig. \ref{fig:fd_dac_delay_high_xi}. Since the FD-DAC and the HD-DAC policy are distinguished when $h_{dac}$ is high, in which case the performance is not influenced by the D2D communication, only the HD-DAC policy is considered in the delay comparison with the MPC policy.

\subsubsection{Delay Comparison between the MPC and the HD-DAC policy}

The MPC policy maximizes the probability of zero delay through cache hits, but the HD-DAC policy may still offer lower delays due to the improvement in the transmission rates. Based on this observation, the two policies are compared in terms of the 90-th percentile of the content retrieval delay, which is an important QoS metric, representing the maximum delay that is experienced by the target UE for 90\% of the time.

\begin{figure}[!t]
\centering
\subfloat[$K=50$]{\includegraphics[width=3in]{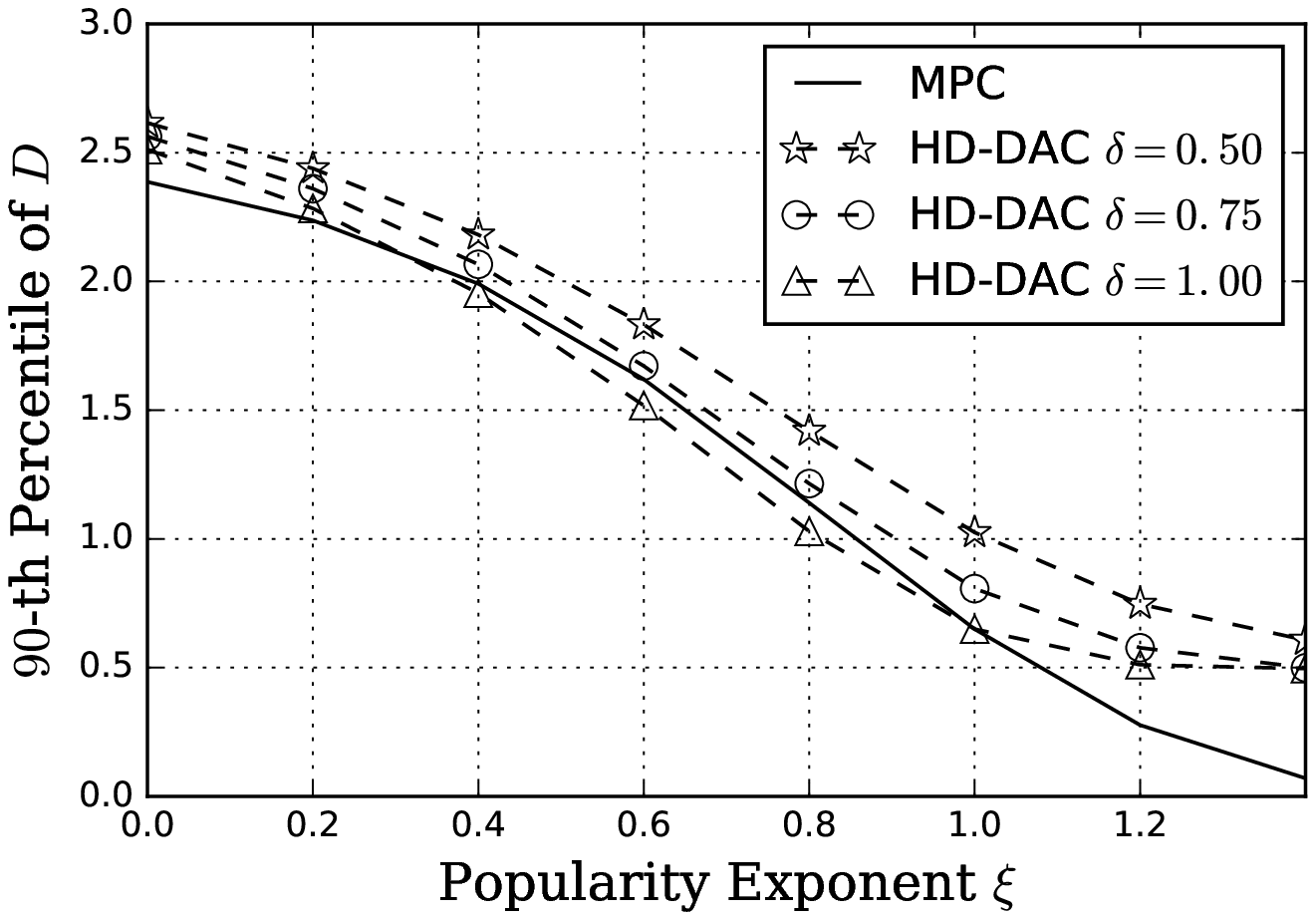}
\label{fig:hd_delay_90th_percentile_K50}
}\hfil
\subfloat[$K=100$]{
\includegraphics[width=3in]{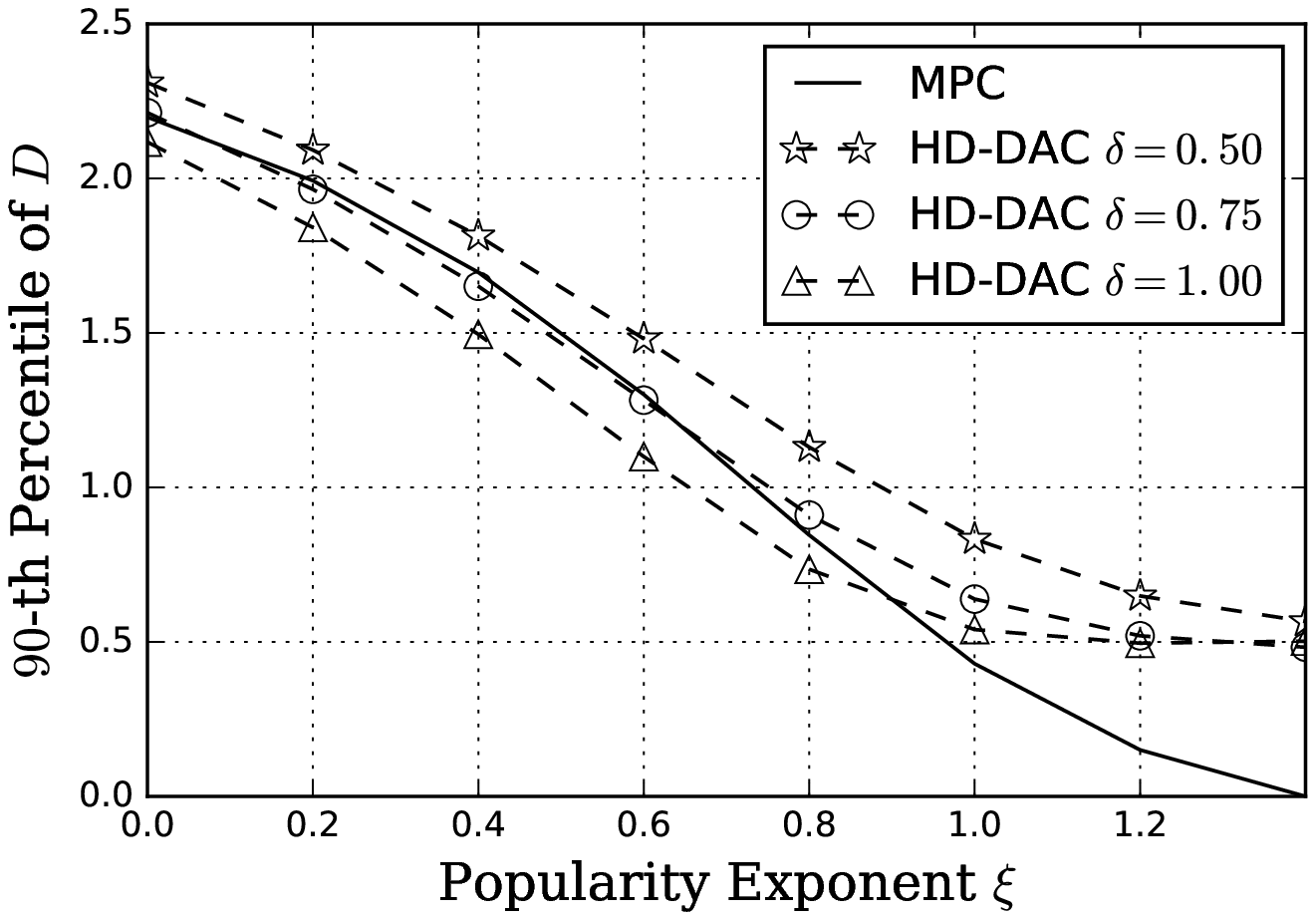}
\label{fig:hd_delay_90th_percentile_K100}
}\hfil
\subfloat[$K=200$]{
\includegraphics[width=3in]{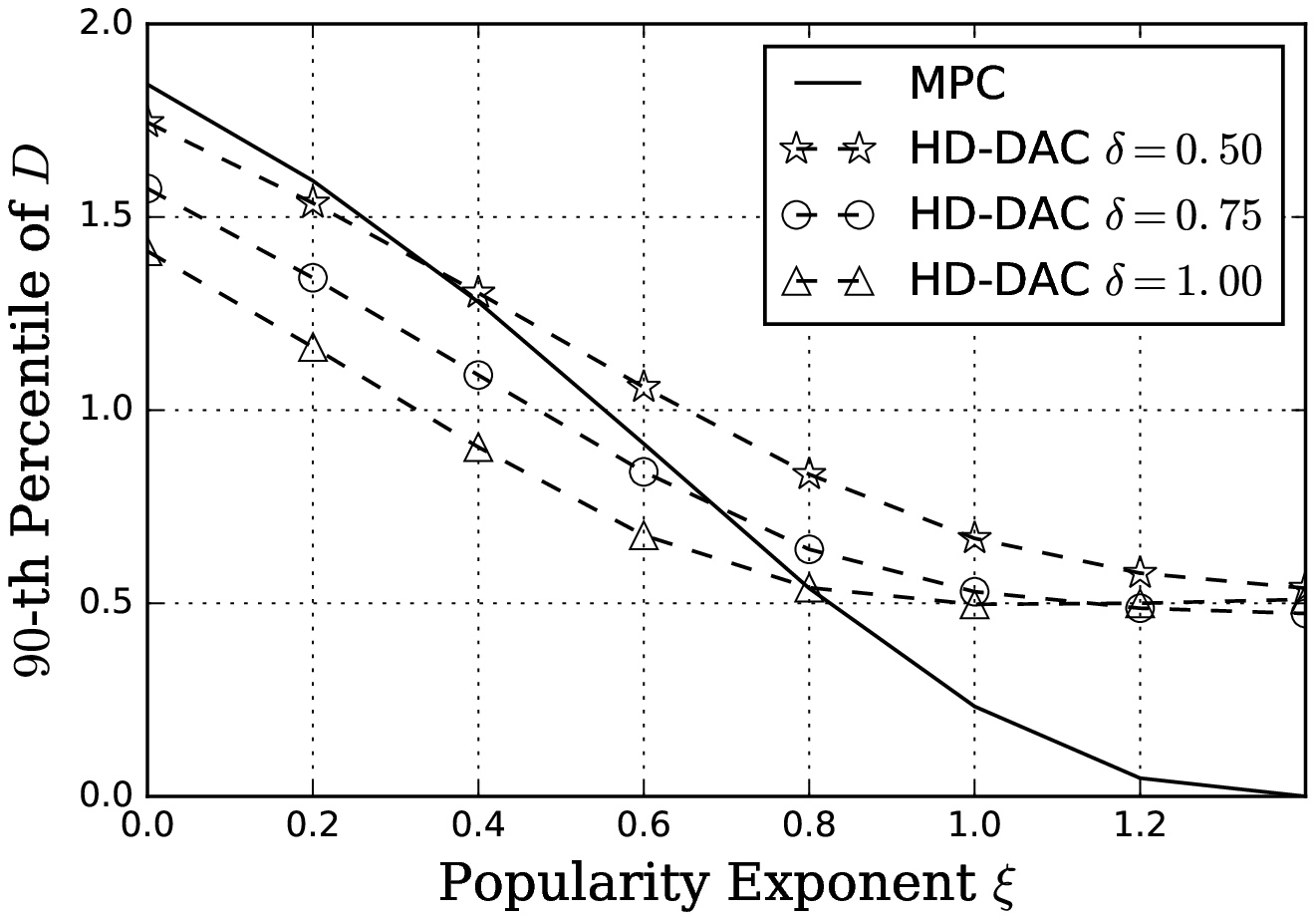}
\label{fig:hd_delay_90th_percentile_K200}
}
\caption{The 90-th percentile of the content retrieval delay $D$ in terms of $\xi$ for a) $K=50$, b) $K=100$, and c) $K=200$.}
\label{fig:hd_delay_90th_percentile}
\end{figure}

In Fig. \ref{fig:hd_delay_90th_percentile}, we plot the delay percentiles for the HD-DAC and the MPC policy as a function of the popularity exponent $\xi$ for the cases: a) $K=50$, b) $K=100$, and c) $K=200$. As a general observation, the 90-th percentile of delay for both policies decreases with higher values of $K$, since both the hit probability and, in the case of the HD-DAC policy, the probability of D2D content exchange, are higher. The delay percentile of the HD-DAC policy also decreases with $\delta$, since the opportunities for D2D communication are improved with a larger number of D2D pairs, while the MPC policy is not affected. In Fig.  \ref{fig:hd_delay_90th_percentile_K50}, the performance is comparable between the HD-DAC policy with $\delta=1.0$, and the MPC policy, for $\xi<1.0$. In Fig. \ref{fig:hd_delay_90th_percentile_K100}, the performance is comparable between the HD-DAC policy with $\delta=0.75$, and the MPC policy, for $\xi<0.8$. In Fig.  \ref{fig:hd_delay_90th_percentile_K200}, the performance is comparable between the HD-DAC policy with $\delta=0.5$, and the MPC policy, for $\xi<0.4$. Based on these these observations, we conclude that, for low values of $\xi$, the HD-DAC policy is favored by larger UE caches and requires fewer D2D pairings to outperform the MPC policy, while for high values of $\xi$, the MPC policy is favored by larger UE caches due to the wide gap in the hit probabilities of the two policies, which justifies the superior performance of the MPC policy in these cases.

\section{Conclusion}\label{section:conclusion}

In this work, we have proposed a novel policy for device caching that combines the emerging technologies of D2D and mmWave communication to enhance the offloading and the delay performance of the cellular network. Based on a stochastic-geometry modeling, we have derived the offloading gain and the distribution of the content retrieval delay for the proposed DAC policy and the state-of-the-art MPC policy, which does not exploit content exchange among the UEs. By comparing analytically and through Monte-Carlo simulations the two policies, we have shown that the proposed policy exhibits superior offloading and delay performance when the availability of pairs in the system is sufficiently high and the popularity distribution of the requested content is not excessively skewed. In addition, motivated by the prospect of bidirectional content exchange, we presented an FD version of the proposed policy, which exhibits a small improvement over the HD version in terms of the delay performance, due to the low probability of bidirectional content exchange. According to the simulation results, increasing this probability does not yield a proportional improvement in performance due to the resulting prevalence of the cellular rate over the D2D rate, attributed to offloading.

As future work, we plan to generalize the proposed caching scheme to a policy that divides the cacheable content to an arbitrary number of groups and study the impact on performance.



\ifCLASSOPTIONcaptionsoff
  \newpage
\fi


\section*{Acknowledgements}
The authors would like to cordially thank the editor and the anonymous reviewers for their constructive suggestions that helped to improve the quality of this work.

\appendices
\section{Proof of Proposition 1}
\label{appendix:proposition_1}
Denoting by A and B the users of a D2D pair and by $\mathcal{C}_A$ and $\mathcal{C}_B$ their caches, the hit probabilities $h_A$ and $h_B$ of the two users can be expressed as
\begin{gather}
h_A = \sum_{i \in \mathcal{C}_A} q_i \nonumber\\
h_B = \sum_{i \in \mathcal{C}_B} q_i, 
\end{gather}
and the exchange probabilities $e_A$ and $e_B$ as
\begin{gather}
e_A = \sum_{i \in {\mathcal{C}_B \cap \overline{\mathcal{C}_A}}} q_i \nonumber\\
e_B = \sum_{i \in {\mathcal{C}_A \cap \overline{\mathcal{C}_B}}} q_i,
\end{gather}
where $\overline{\mbox{} \cdot \mbox{}}$ signifies the complement in terms of the set of the library contents.

To prove that $e_A$ and $e_B$ are maximized when $\mathcal{C}_A$ and $\mathcal{C}_B$ form a partition of the $2K$ most popular contents, we need to show that i) the optimal $\mathcal{C}_A$ and $\mathcal{C}_B$ do not overlap, i.e.,  $\mathcal{C}_A \cap \mathcal{C}_B = \emptyset$, and ii) the optimal $\mathcal{C}_A$ and $\mathcal{C}_B$ cover the $2K$ most popular contents, i.e., $\mathcal{C}_A \cup \mathcal{C}_B = \{i \in \mathbb{N}: 1 \leq i \leq 2 K \}$. We prove both i) and ii) by contradiction. Regarding i), if the optimal $\mathcal{C}_A$ and $\mathcal{C}_B$ contained a common content, say $c \in \mathcal{C}_A \cap \mathcal{C}_B $, we could simultaneously increase $e_A$ and $e_B$ by replacing $c$ in $\mathcal{C}_A$ with a content from $\overline{\mathcal{C}_A} \cap \overline{\mathcal{C}_B}$. Therefore, $\mathcal{C}_A$ and $\mathcal{C}_B$ must not overlap. Regarding ii), if $\mathcal{C}_A$ contained a content $c$ that did not belong in the $2K$ most popular, then we could replace $c$ with an uncached content from $\{i \in \mathbb{N}: 1 \leq i \leq 2 K \}$, which would increase $e_B$ and $h_A$, while leaving $e_A$ and $h_B$ unaffected. Therefore, if $\mathcal{C}_A$ and $\mathcal{C}_B$ form a partition of $\{i \in \mathbb{N}: 1 \leq i \leq 2 K \}$, neither $e_A$, $e_B$ nor $h_A$, $h_B$ can be increased simultaneously with a different partition.

\section{Proof of Proposition 2}
\label{appendix:hitprob_analysis}
The ratio of the hit probabilities of the two policies, $h_{ratio}$, is given by \eqref{hratio}, which we repeat here for easier reference:
\begin{equation}
h_{ratio} = \frac{h_{dac}}{h_{mpc}} = \frac{1}{2} \frac{\sum_{i=1}^{2K} i^{-\xi}}{\sum_{j=1}^{K} j^{-\xi}}.
\label{eq:hratio_def}
\end{equation}

To prove that $h_{ratio}$ decreases monotonically with $\xi$, we differentiate $h_{ratio}$ in terms of $\xi$ as
\begin{align}
\frac{\partial h_{ratio}}{\partial \xi} & =
-\frac{1} {2 \left(\sum_{j=1}^{K} j^{-\xi}\right)^2} 
\sum_{i=1}^{2K} \sum_{j=1}^{K} (ij)^{-\xi} \left[ \ln (i) - \ln(j) \right] = \nonumber\\ 
& = -\frac{1} {2 \left( \sum_{j=1}^{K} j^{-\xi}\right)^2} 
\left[
\sum_{i=1}^{K} \sum_{j=1}^{K} (ij)^{-\xi} \left[ \ln (i) - \ln(j) \right] + \sum_{i=K+1}^{2K} \sum_{j=1}^{K} (ij)^{-\xi} \left[ \ln (i) - \ln( j) \right]  
\right] \overset{(a)}= \nonumber\\
& = -\frac{1} {2 \left(\sum_{j=1}^{K} j^{-\xi}\right)^2} 
\sum_{i=K+1}^{2K} \sum_{j=1}^{K} (ij)^{-\xi} \left[ \ln (i) - \ln(j) \right]  \overset{(b)}< 0,
\end{align}
where $(a)$ follows because the first sum is eliminated due to symmetry, and $(b)$ follows because $\ln(i)>\ln(j)$ for the remaining indexes. Since the derivative of $h_{ratio}$ in terms of $\xi$ is negative, $h_{ratio}$ decreases monotonically with $\xi$.

To prove that $h_{ratio}$ decreases monotonically with $K$, we need to show that $h_{ratio}(K+1) < h_{ratio}(K)$. Introducing the notation $\mathsf{S}_K \triangleq \sum_{i=1}^{K} i^{-\xi}$ for clarity, the aforementioned inequality is transformed as
\begin{equation}
h_{ratio}(K+1) < h_{ratio}(K)
\, \Leftrightarrow \,
\frac{1}{2}\frac{\mathsf{S}_{2K+2}}{\mathsf{S}_{K+1}} < \frac{1}{2} \frac{\mathsf{S}_{2K}}{\mathsf{S}_K} 
\, \Leftrightarrow \,
\frac{\mathsf{S}_{2K} + (2K+1)^{-\xi} + (2K+2)^{-\xi}}{\mathsf{S}_{K} + (K+1)^{-\xi}} < \frac{\mathsf{S}_{2K}}{\mathsf{S}_K}
\end{equation}
Manipulating the inequality yields
\begin{equation}
\frac{\mathsf{S}_{2K}}{\mathsf{S}_{K}} > \left( \frac{2 K+1}{K+1} \right)^{-\xi} + 2^{-\xi}.
\end{equation}
Splitting the odd and even indexes in $\mathsf{S}_{2K}$ as
\begin{equation}
\mathsf{S}_{2K} = \sum_{i=1}^{K} (2 i-1)^{-\xi} + \sum_{i=1}^{K} (2 i)^{-\xi}
 = \sum_{i=1}^{K} (2 i-1)^{-\xi} + 2^{-\xi} \mathsf{S}_{K},
\end{equation}
the inequality is further simplified to
\begin{equation}
\frac{ \sum_{i=1}^{K} (2 i-1)^{-\xi} }{ \sum_{i=1}^{K} i^{-\xi} } > \left( \frac{2 K+1}{K+1} \right)^{-\xi} 
\, \Leftrightarrow \,
\sum_{i=1}^{K} \left( \frac{2 i-1}{2 K+1} \right)^{-\xi} >
\sum_{i=1}^{K} \left( \frac{i}{ K+1} \right)^{-\xi}
\end{equation}
Comparing the sums term-by-term, the inequality holds provided that
\begin{equation}
\frac{2 i-1}{2 K+1} < \frac{i}{K+1} \, \Leftrightarrow \, i<K+1.
\end{equation}
Since the final inequality is true and all the steps of the derivation were reversible, the initial inequality is also proven.

To calculate the limit of $h_{ratio}$ for high values of $K$, we distinguish the cases $\xi >1$ and $\xi \leq 1$.
\begin{itemize}
\item For  $\xi>1$, the sums in \eqref{eq:hratio_def} converge as $K \to \infty$, yielding
\begin{equation}
\lim_{K \to \infty} h_{ratio} = \frac{1}{2} \frac{\zeta(\xi)}{\zeta(\xi)} = \frac{1}{2},
\label{eq:hratio_lim_high_xi}
\end{equation}
where $\zeta(\cdot)$ is the Rieman-Zeta function.

\item For $\xi \leq 1$, the sums in \eqref{eq:hratio_def} diverge as $K \to \infty$, nevertheless, the limit can be calculated through an asymptotic expression of the sums, based on the Euler-McLaurin summation formula \cite{general:analytic_number_theory}. According to this formula, the discrete sum can be approximated with a continuous integral as 
\begin{equation}
\sum_{i=1}^{K} i^{-\xi} \sim  \int_{1}^K i^{-\xi} di + \epsilon(\xi) = 
\begin{cases} \frac{K^{1-\xi}-1}{1-\xi} + \epsilon(\xi) & \mbox{if } \xi < 1 \\
\ln (K) + \epsilon(1) & \mbox{if } \xi = 1
 \end{cases}, \label{eq:euler-mclaurin}
\end{equation}
where $\epsilon(\xi)$ represents the asymptotic error of the approximation, also known as the \textit{generalized Euler constant}\footnote{For the special case $\xi=1$,$\epsilon(1)$ is the well-known Euler–Mascheroni constant, equal to 0.57721..}. Applying \eqref{eq:euler-mclaurin} to \eqref{eq:hratio_def} yields
\begin{equation}
\lim_{K \to \infty} h_{ratio} =
\begin{cases}
\lim_{K \to \infty} \frac{1}{2} \frac{\frac{(2K)^{1-\xi}-1}{1-\xi} + \epsilon(\xi)} {\frac{K^{1-\xi}-1}{1-\xi} + \epsilon(\xi)} =
2^{-\xi} & \mbox{if } \xi < 1 \\
\lim_{K \to \infty} \frac{1}{2} \frac{\ln (2K) + \epsilon(1)} {\ln (K) + \epsilon(1)} =
\frac{1}{2} & \mbox{if } \xi = 1 \end{cases} . \label{eq:hratio_lim_small_xi}
\end{equation}
\end{itemize} 

Finally, the limits in \eqref{eq:hratio_lim_high_xi} and \eqref{eq:hratio_lim_small_xi} can be combined in one compact expression as
\begin{equation}
\lim_{K \to \infty} h_{ratio} = \mbox{max}\left(2^{-\xi}, \frac{1}{2}\right), \mbox{ } \xi \geq 0
\end{equation}

\section{Derivation of the CCDF of the cellular SINR}
\label{appendix:cellular_SINR}
Defining
\begin{equation}
J_1(T,a) \triangleq \int_0^{\infty} \frac{2r}{r_{cell}^2}e^{-\left(\frac{r}{r_{cell}}\right)^2} e^{-\hat{N}Tr^{a}}dr
\label{eq:J1}
\end{equation}
and
\begin{equation}
J_2(T,a) \triangleq \int_0^{\infty} \frac{2r}{r_{cell}^2}
e^{-\left(\frac{r}{r_{cell}}\right)^2}
e^{-\frac{r}{r_{los}}}
e^{-\hat{N}Tr^a} dr,
\label{eq:J2}
\end{equation}
the CCDF of the cellular SINR is expressed as
\begin{equation}
\text{P}(SINR_{cell}>T) \approx \text{P}(SNR_{cell}>T) = J_1(T,a_{N})+J_2(T,a_{L})-J_2(T,a_N).
\label{eq:cell_sinr_appendix}
\end{equation}
Since the integrals in \eqref{eq:J1} and \eqref{eq:J2} cannot be evaluated in closed form, they are approximated as follows.
\begin{itemize}
\item Regarding $J_1$, the exponential term $e^{-\left(\frac{r}{r_{cell}}\right)^2}$ is approximated with a piecewise linear function as
\begin{equation}
e^{-\left(\frac{r}{r_{cell}}\right)^2} \approx
\left\{
	\begin{array}{ll}
		1-\frac{r}{r_1} & \text{for }r \leq r_1 \\
		0 & \text{for }r>r_1
	\end{array}
\right.,
\end{equation}
yielding
\begin{equation}
J_1(T,a) \approx \int_0^{r_1} \frac{2r}{r_{cell}^2}\left(1-\frac{r}{r_1}\right) e^{-\hat{N}Tr^{a}}dr=
\frac{2}{a r_{cell}^2} \left( 
\frac{\gamma\left(\frac{2}{a}, \hat{N} T r_1^{a}\right)}{(\hat{N}T)^{\frac{2}{a}}} -
\frac{\gamma\left(\frac{3}{a}, \hat{N} T r_1^{a}\right)}{r_1 (\hat{N}T)^{\frac{3}{a}}} \right) 
\label{eq:J1_approx},
\end{equation}
where $r_1$ is chosen so that the approximated value of $J_1$ is exact for $T=0$, i.e.,
\begin{equation}
J_1(0,a)=\int_0^{\infty} \frac{2r}{r_{cell}^2}e^{-\left(\frac{r}{r_{cell}}\right)^2}dr =
\int_0^{r_1} \frac{2r}{r_{cell}^2}\left(1-\frac{r}{r_1}\right)dr \mbox{ }\Rightarrow\mbox{ } r_1= \sqrt{3}r_{cell}.\label{eq:r1_derivation}
\end{equation}

\item Regarding $J_2$, the exponential term $e^{-\left(\frac{r}{r_{cell}}\right)^2}e^{-\frac{r}{r_{los}}}$ is approximated with a quadratic function as
\begin{equation}
e^{-\left(\frac{r}{r_{cell}}\right)^2}e^{-\frac{r}{r_{los}}} \approx
\left\{
	\begin{array}{ll}
		\left(1-\frac{r}{r_{2}}\right)^2 & \text{for }r \leq r_{2} \\
		0 & \text{for }r>r_{2}
	\end{array}
\right.,
\end{equation}
yielding
\begin{align}
&J_2(T,a)\approx \int_0^{\infty} \frac{2r}{r_{cell}^2} \left(1-\frac{r}{r_{2}}\right)^2 e^{-\hat{N}Tr^a} dr = \\&= \frac{2}{a r_{cell}^2} \left( 
\frac{\gamma\left(\frac{2}{a}, \hat{N} T r_2^{a}\right)}{(\hat{N}T)^{\frac{2}{a}}} -
2\frac{\gamma\left(\frac{3}{a}, \hat{N} T r_2^{a}\right)}{r_2 (\hat{N}T)^{\frac{3}{a}}} +
\frac{\gamma\left(\frac{4}{a}, \hat{N} T r_2^{a}\right)}{r_2^2 (\hat{N}T)^{\frac{4}{a}}}\right), \label{eq:J2_approx}
\end{align}
where $r_2$ is chosen so that the approximated value of $J_2$ is exact for $T=0$, i.e.,
\begin{align}
J_2(0,a)&= \int_0^{\infty} \frac{2r}{r_{cell}^2}
e^{-\left(\frac{r}{r_{cell}}\right)^2}
e^{-\frac{r}{r_{los}}} dr = 
\int_0^{r_2} \frac{2r}{r_{cell}^2}\left(1-\frac{r}{r_{2}}\right)^2dr
\nonumber\\ & \Rightarrow \mbox{ }
r_2=\sqrt{6}  \sqrt{1-\sqrt{\pi}\frac{r_{cell}}{2 r_{los}} e^{\left(\frac{r_{cell}}{2 r_{los}}\right)^2} \text{erfc}\left(\frac{r_{cell}}{2r_{los}}\right)} r_{cell}.\label{eq:r2_derivation}
\end{align}
\end{itemize} 

Combining \eqref{eq:J1_approx} and \eqref{eq:J2_approx} into \eqref{eq:cell_sinr_appendix} yields the final result.

\section{Derivation of the Laplace transform of the HD D2D interference}
\label{appendix:hd-dac}
The D2D interference in the HD-DAC policy is given by
\begin{equation}
\hat{I} = \sum_{x \in \Phi_{d2d}^{hd}} g_x \eta_x  r_x^{-a_x}
\label{eq:appD:interf_definition}
\end{equation}
where $\Phi_{d2d}^{hd}$ is the PPP of the D2D interferers, with intensity 
\begin{equation}
\lambda_{d2d}^{hd}=\left(1-(1-h_{dac})^2 \right) \lambda_{p}=\frac{\delta}{2} h_{dac}\left(2-h_{dac} \right) \lambda_{ue}.
\end{equation}
Based on \eqref{eq:appD:interf_definition}, the Laplace transform of the HD D2D interference, denoted by $\mathcal{L}_{\hat{I}}^{hd}(s)$, is derived as
\begin{align}
&\mathcal{L}_{\hat{I}}^{hd}(s)=\mathbb{E}[e^{-\hat{I} s}] = 
\mathbb{E}_{\Phi_{d2d}^{hd}} 
\left[\Pi_{x \in \Phi_{d2d}^{hd}} \mathbb{E}\left[e^{- g_x \eta_x  r_x^{-a_x} s}\right]\right]\overset{(i)}= \nonumber\\
=&e^{-\lambda_{d2d}^{hd}\int_0^{2\pi} \int_0^{\infty}\left(1-\mathbb{E}\left[e^{- g  \eta  r^{-a} s}\right]\right) r dr d\phi } \overset{(ii)}=
e^{-\pi \delta h_{dac}\left(2-h_{dac} \right) \lambda_{ue} \int_0^{\infty} \left(1-\mathbb{E} \left[\frac{1}{1+ g  r^{-a} s} \right]\right) r dr },
\label{eq:appD:interf_laplace}
\end{align}
where $(i)$ follows from the probability generating functional (PGFL) of the PPP \cite{stochgeom:haenggi_book}, and $(ii)$ from the Laplace transform of the exponential random variable.

Defining
\begin{align}
J_3(s,a) \triangleq  \int_0^{\infty} \frac{g   r^{-a} s}{1+g r^{-a} s} r dr
\end{align}
and
\begin{align}
J_4\left(s,a \right) \triangleq \int_0^{\infty}  \frac{e^{-\frac{r}{r_{los}}}}{1+g r^{-a} s} r dr,
\end{align}
the integral in the exponent of \eqref{eq:appD:interf_laplace} can be expressed as
\begin{align}
& \int_0^{\infty} \left(1-\mathbb{E} \left[\frac{1}{1+ g  r^{-a} s} \right]\right) r dr =
\mathbb{E}_g \left[ \int_0^{\infty}
\left(
1 - 
\frac{e^{-\frac{r}{r_{los}}}}{1+ g r^{-a_L} s}
- \frac{1-e^{-\frac{r}{r_{los}}}}{1+ g r^{-a_N} s} 
\right) r dr
\right] = \nonumber\\
= & 
\mathbb{E}_g \left[
\int_0^{\infty} \left( 1 - \frac{1}{1+ g r^{-a_N} s} \right) r dr +
\int_0^{\infty} \frac{e^{-\frac{r}{r_{los}}}}{1+ g r^{-a_N} s} r dr -
\int_0^{\infty} \frac{e^{-\frac{r}{r_{los}}}}{1+ g r^{-a_L} s} r dr
\right] = \nonumber\\
= & \mathbb{E}_{g} \left[ J_3(s,a_N) + J_4(s,a_N)-J_4(s,a_L)\right]
\label{eq:appD:integral_in_exponent}
\end{align}
Subsequently, $J_3(s,a)$ is evaluated in closed form as
\begin{equation}
J_3(s,a) = \frac{1}{2}
\Gamma \left(1-\frac{2}{a}\right)
\Gamma \left(1+\frac{2}{a}\right)
(gs)^{\frac{2}{a}},
\label{eq:appD:psi3}
\end{equation}
while $J_4(s,a)$ is derived through the approximation
\begin{equation}
e^{-\frac{r}{r_{los}}} \approx \left(1-\frac{r}{r_4}\right)^k,
\label{eq:appD:approximation}
\end{equation}
yielding
\begin{align}
J_4\left(s,a \right) \approx \int_0^{r_4} \frac{
\left(1-\frac{r}{r_4}\right)^k
}{1+g r^{-a} s} r dr = 
\sum_{l=0}^k {k \choose l}  \frac{(-1)^l}{r_4^l}
\int_0^{r_4}  \frac{
 r^{l+1}
}{1+g r^{-a} s} dr = \nonumber\\
\sum_{l=0}^k {k \choose l} (-1)^l \frac{r_4^{a+2}}{(l+a+2)gs}
{}_2 F_1\left(1,1+\frac{l+2}{a};2+\frac{l+2}{a};-\frac{r_4^a}{gs}\right),
\label{eq:appD:psi4}
\end{align}
where $r_4$ is chosen so that the approximated value of $J_4$ is exact for $s=0$, i.e.,
\begin{equation}
J_4(0,a) = \int_0^{\infty} e^{-\frac{r}{r_{los}}} r dr 
= \int_0^{r_4} \left(1-\frac{r}{r_4}\right)^k r dr
\mbox{ } \Rightarrow \mbox{ }
r_4 = \sqrt{(k+1)(k+2)} r_{los}.
\label{eq:appD:approximation_r4}
\end{equation}
Applying \eqref{eq:appD:approximation_r4} to \eqref{eq:appD:approximation}, we also observe that the approximation becomes exact as $k$ grows, since
\begin{equation}
\lim_{k \to \infty} \left(1-\frac{r}{\sqrt{(k+1)(k+2)} r_{los}}\right)^k = \lim_{k \to \infty} \left(1-\frac{r}{k r_{los}}\right)^k = e^{-\frac{r}{r_{los}}}.
\end{equation}

Finally, combining \eqref{eq:appD:psi3} and \eqref{eq:appD:psi4} into \eqref{eq:appD:integral_in_exponent} and the result to \eqref{eq:appD:interf_laplace} yields the final result. Please note that the remaining expectation over $g$ is trivial, since $g$ is a discrete random variable with the distribution
\begin{equation}
g=
\left\{
	\begin{array}{ll}
		1 & \mbox{with probability } \frac{\Delta\theta_{ue}^2}{4\pi^2} \\
		\frac{G_{ue}^{min}}{G_{ue}^{max}}  & \mbox{with probability } \frac{2 \Delta\theta_{ue}(2\pi-\Delta\theta_{ue})}{4\pi^2} \\
		\left(\frac{G_{ue}^{min}}{G_{ue}^{max}}\right)^2 & \mbox{with probability } \frac{(2\pi-\Delta\theta_{ue})^2}{4\pi^2}
	\end{array}
\right..
\label{def:normed_link_gain_as_rv}
\end{equation}

\section{Derivation of the Laplace transform of the FD interference}
\label{appendix:fd-dac}

The D2D interference in the FD-DAC policy is given by 
\begin{equation}
\hat{I} = \sum_{x \in \Phi_{d2d}^{fd(1)}} g_x \eta_x  r_x^{-a_x} + \sum_{y \in \Phi_{d2d}^{fd(2)}} g_y \eta_y  r_y^{-a_y},
\label{eq:appE:fd_interf_definition}
\end{equation}
where $\Phi_{d2d}^{fd(1)}$ and $\Phi_{d2d}^{fd(2)}$ are the PPPs of the D2D interferers, both with intensity
\begin{equation}
\lambda_{d2d}^{fd}=h_{dac} \lambda_{p}=\frac{\delta}{2} h_{dac} \lambda_{ue},
\end{equation}
but dependent to each other due to the D2D pairings. Based on \eqref{eq:appE:fd_interf_definition}, the Laplace tranform of the D2D interference in the FD-DAC policy, denoted by $\mathcal{L}_{\hat{I}}^{fd}(s)$, is expressed as
\begin{align}
&\mathcal{L}_{\hat{I}}^{fd}(s)=
\mathbb{E} \left[e^{-\hat{I} s}\right] = 
\mathbb{E} \left[
\Pi_{x \in \Phi_{d2d}^{fd(1)}} 
 e^{- g_x \eta_x  r_x^{-a_x} s}\cdot
\Pi_{y \in \Phi_{d2d}^{fd(2)}} 
 e^{- g_y \eta_y  r_y^{-a_y} s}
\right].
\label{eq:appE:fd_interf_laplace}
\end{align}

Due to the dependence of $\Phi_{d2d}^{fd(1)}$ and $\Phi_{d2d}^{fd(2)}$, \eqref{eq:appE:fd_interf_laplace} cannot be evaluated in closed form, nevertheless, it can be approximated with the following bounds\cite{fd:stoch_geom_anal}.
\begin{itemize}
\item From the FKG inequality
\begin{align}
&\mathcal{L}_{\hat{I}}^{fd}(s) \geq 
\mathbb{E} \left[
\Pi_{x \in \Phi_{d2d}^{fd(1)}} 
 e^{- g_x \eta_x  r_x^{-a_x} s}
\right] \cdot
\mathbb{E} \left[
\Pi_{y \in \Phi_{d2d}^{fd(2)}} 
 e^{- g_y \eta_y  r_y^{-a_y} s} \right]
= \nonumber\\ & = 
 e^{- \pi \delta \lambda_{ue} h_{dac} 2  \mathbb{E}_g
 \left[
 J_3\left(s,a_N \right) + J_4\left(s,a_N \right) - J_4\left(s,a_L\right)
 \right]}
\label{eq:appE:fd_interf_laplace_fkg_bound}
\end{align}

\item From the Cauchy-Schwarz inequality,
\begin{align}
&\mathcal{L}_{\hat{I}}^{fd}(s) \leq
\sqrt{
\mathbb{E} \left[
\Pi_{x \in \Phi_{d2d}^{fd(1)}} 
 e^{- 2 g_x \eta_x  r_x^{-a_x} s}
\right] \cdot
\mathbb{E} \left[
\Pi_{y \in \Phi_{d2d}^{fd(2)}} 
 e^{- 2 g_y \eta_y  r_y^{-a_y} s} \right]
 }
 = \nonumber\\& = e^{-\pi \delta \lambda_{ue} h_{dac} \mathbb{E}_g
 \left[
 J_3\left(2s,a_N \right) + J_4\left(2s,a_N\right) - J_4\left(2s,a_L\right)
 \right]}
\label{eq:appE:fd_interf_laplace_cs_bound}
\end{align}

\end{itemize}
In \eqref{eq:appE:fd_interf_laplace_fkg_bound} and \eqref{eq:appE:fd_interf_laplace_cs_bound}, the functions $J_3(s,a)$ and $J_4(s,a)$ are given by \eqref{eq:appD:psi3} and \eqref{eq:appD:psi4} respectively. 


\bibliographystyle{IEEEtran}
\bibliography{IEEEabrv,jsac_refs}

%

\begin{IEEEbiography}
[{\includegraphics[width=1in,height=1.25in,clip,keepaspectratio]{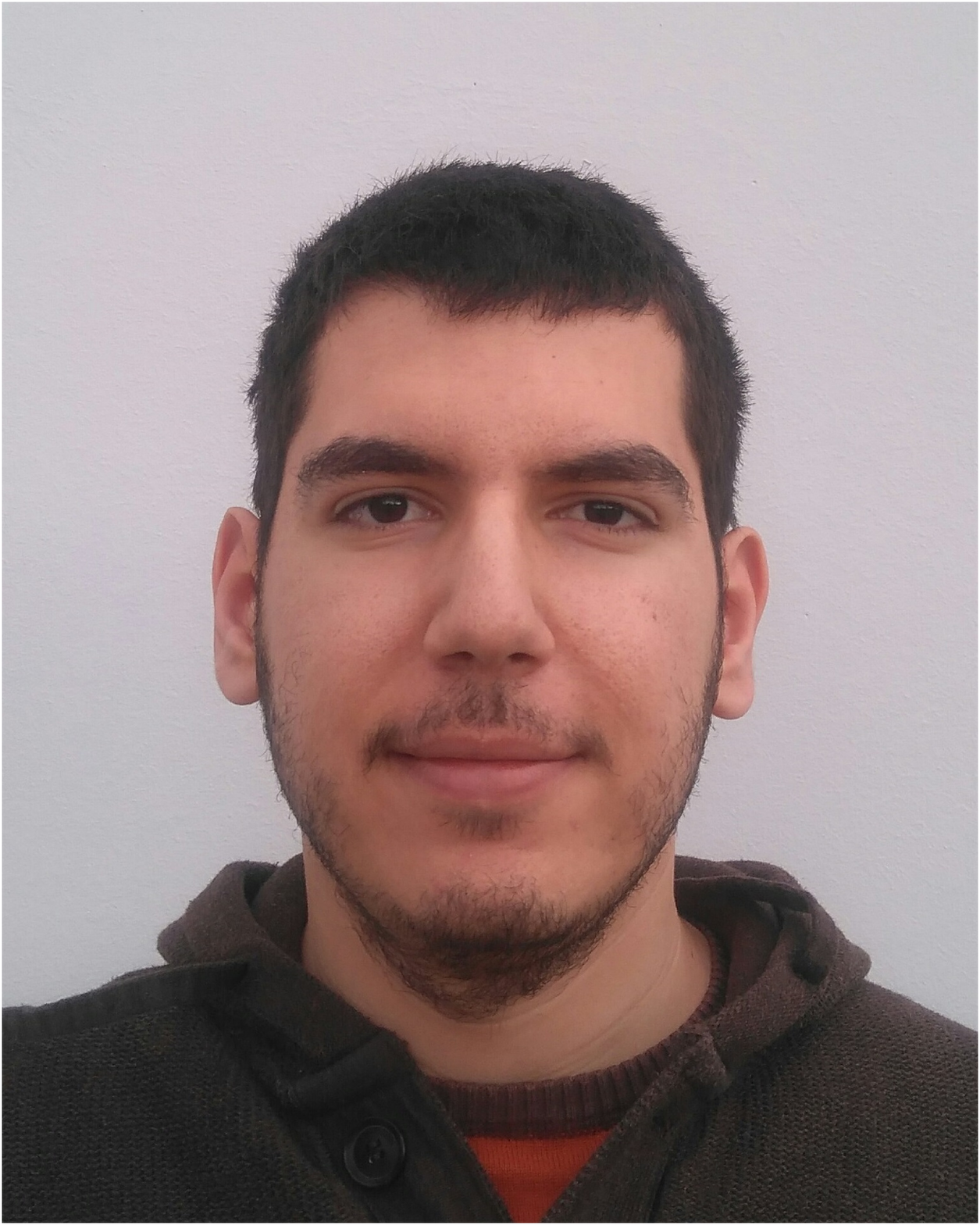}}]
{Nikolaos Giatsoglou} received the Diploma in Electrical and Computer Engineering from the Aristotle University of Thessaloniki (AUTh), Greece, in 2015. He is currently a Ph.D. candidate at the Technical University of Catalonia (UPC) and a Marie-Curie Early Stage Researcher (ESR) working at IQUADRAT Informatica S.L., Barcelona, Spain, in the context of the European project 5Gwireless. His research interests include the performance analysis of wireless protocols, with an emphasis on caching, mmWave, and wireless full-duplex technology.

\end{IEEEbiography}

\begin{IEEEbiography}
[{\includegraphics[width=1in,height=1.25in,clip,keepaspectratio]{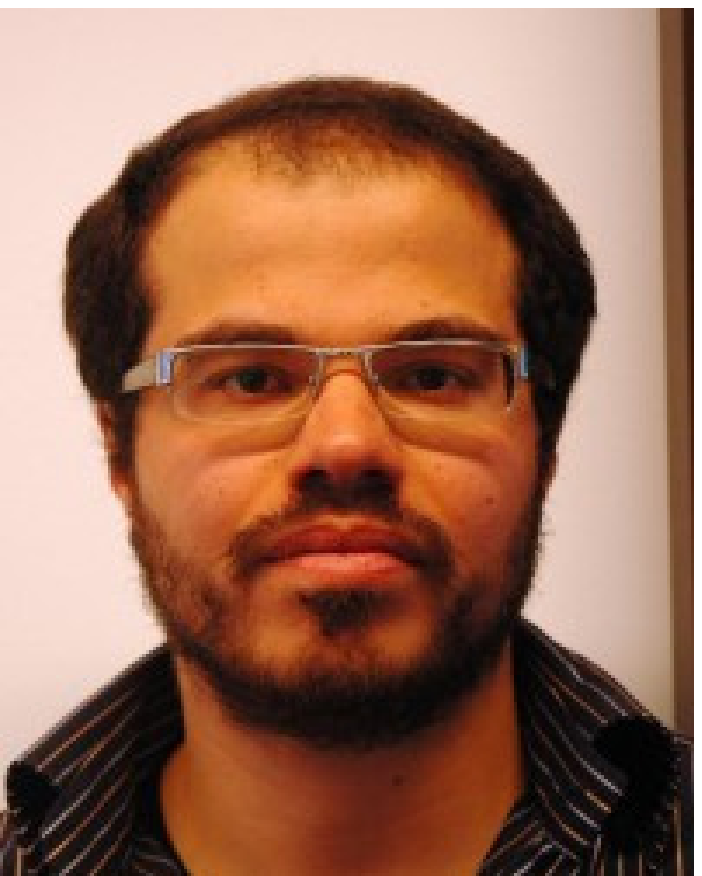}}]
{Konstantinos Ntontin} received the Diploma in Electrical and Computer Engineering in 2006, the M.Sc. Degree in Wireless Systems in 2009, and the Ph.D. degree in 2015 from the University of Patras, Greece, the Royal Institute of Technology (KTH), Sweden, and the Technical University of Catalonia (UPC), Spain, respectively. He is the recipient of the 2013 IEEE COMMUNICATIONS LETTERS Exemplary Reviewer Certificate. His research interests are related to the Physical Layer of wireless telecommunications with an emphasis on the performance analysis in fading channels, MIMO systems, array beamforming, and stochastic modeling of wireless channels.
\end{IEEEbiography}

\begin{IEEEbiography}
[{\includegraphics[width=1in,height=1.25in,clip,keepaspectratio]{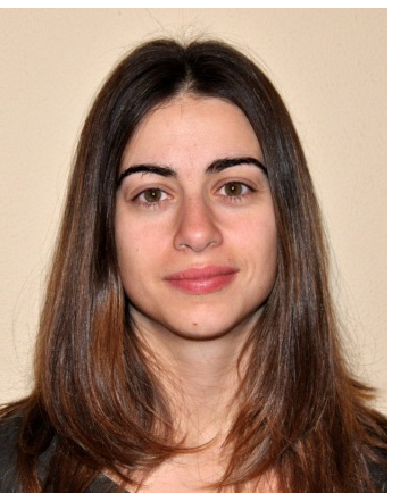}}] 
{Elli Kartsakli} received her Ph.D. degree from the Technical University of Catalonia (UPC) in February 2012 and is currently a senior research engineer in the IQUADRAT R\&D team. Her work has been published and presented in multiple journals, magazines, book chapters and international conferences, and she has actively participated in several national and European projects (IAPP-WSN4QoL, ITN-GREENET, IAPP-Coolness, etc.). Her primary research interests include wireless networking, channel access protocols, energy efficient communication protocols, and protocols and architectures for 5G networks and beyond. 
\end{IEEEbiography}

\begin{IEEEbiography}
[{\includegraphics[width=1in,height=1.25in,clip,keepaspectratio]{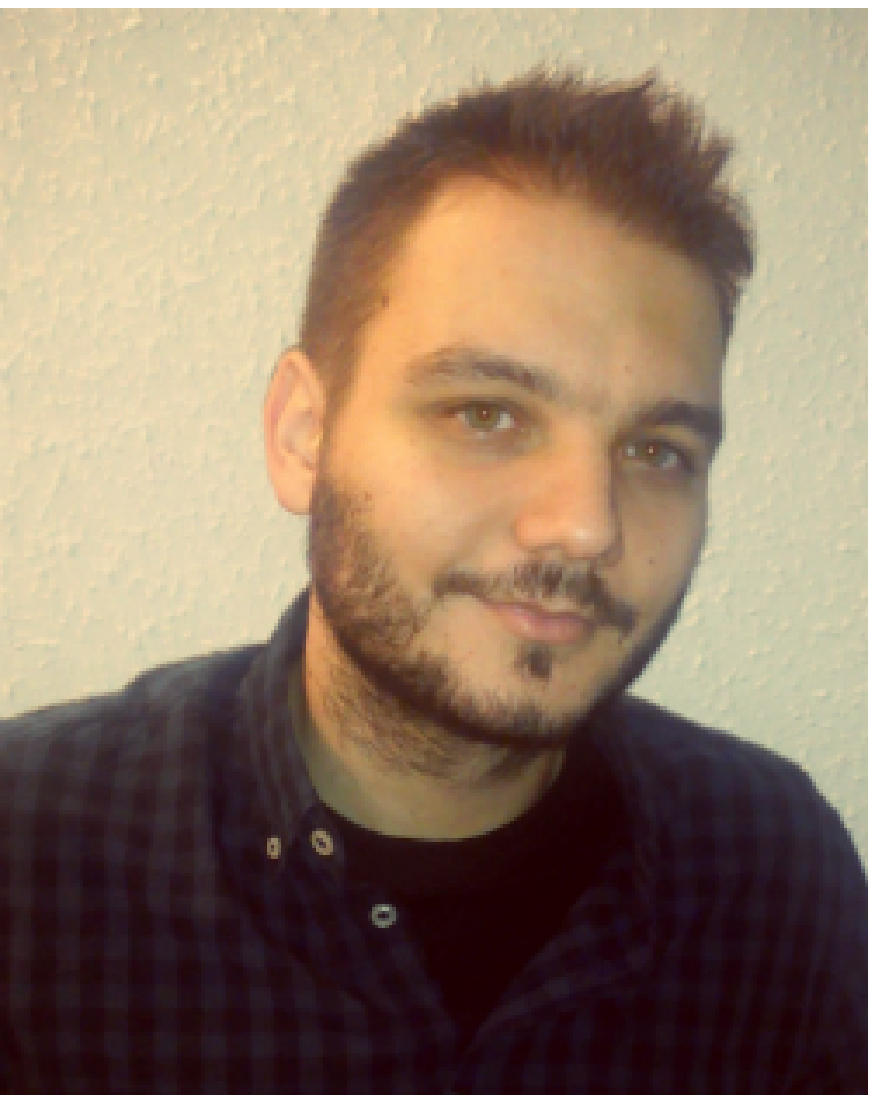}}]
{Angelos Antonopoulos} received his Ph.D. degree from the Technical University of  Catalonia (UPC) in 2012. He is currently a Researcher in the SMARTECH department of the Technological Telecommunications Centre of Catalonia (CTTC). He has published over 70 peer-reviewed journals, conferences and book chapters on various topics, including energy efficient network planning and sharing, 5G wireless networks, cooperative communications and network economics. He has been nominated as Exemplary Reviewer for the IEEE Communications Letters, and has  received the best paper award in IEEE GLOBECOM 2014, the best demo award in IEEE CAMAD 2014, the 1st prize in the IEEE ComSoc Student Competition (as a Mentor) and the EURACON best student paper award in EuCNC 2016. 
\end{IEEEbiography}

\begin{IEEEbiography}
[{\includegraphics[width=1in,height=1.25in,clip,keepaspectratio]{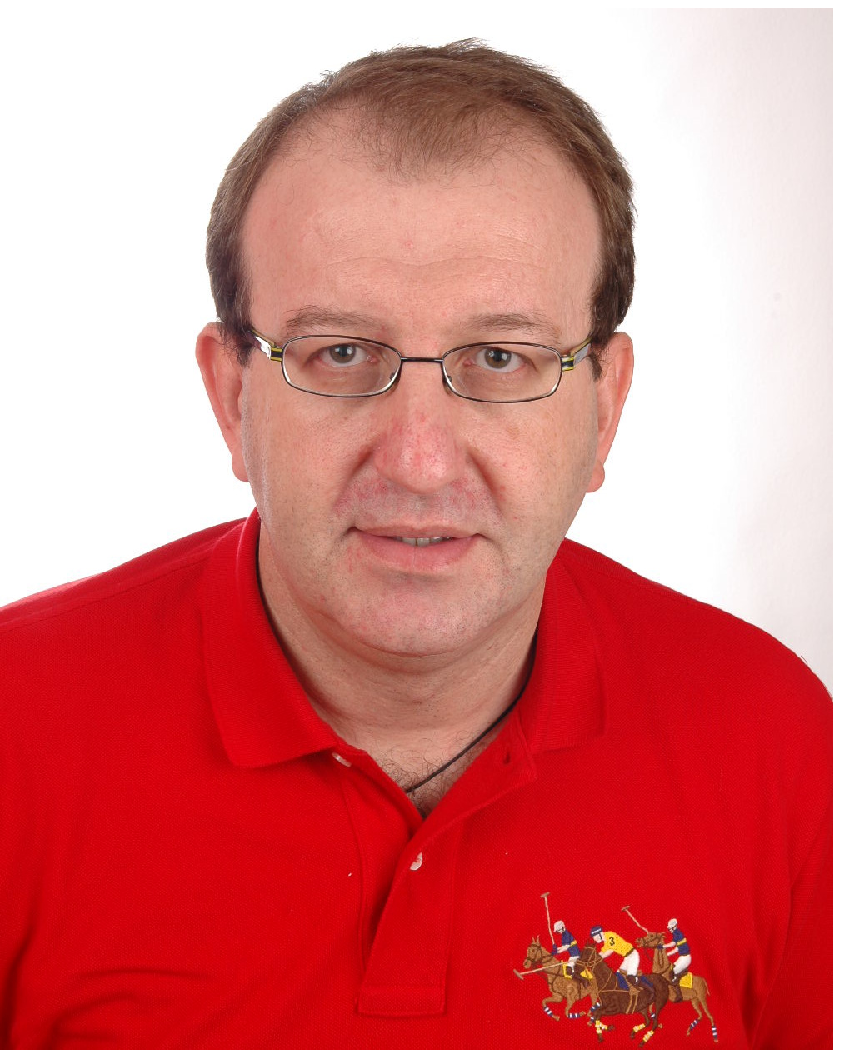}}]
 {Christos Verikoukis} received his Ph.D. degree from the Technical University of Catalonia (UPC) in 2000. He is currently a Fellow Researcher at CTTC (Head of the SMARTECH department) and an adjunct professor at the Electronics Department of the University of Barcelona (UB). He has published 107 journal papers and over 170 conference papers. He is co-author in 3 books, 16 chapters in different books, and has filed 3 patents. He has supervised 15 Ph.D. students and 5 Post Docs researchers since 2004. He has participated in more than 30 competitive projects and has served as the Principal investigator in national projects in Greece and Spain. He received the best paper award in the IEEE ICC 2011, IEEE ICC 2014, IEEE GLOBECOM 2015, and in the EUCNC 2016 conferences, as well as the EURASIP 2013 Best Paper Award for the Journal on Advances in Signal Processing. He was the general Chair of the 17th, 18th, and 19th IEEE CAMAD, the TPC Co-Chair of the 15th IEEE Healthcom and the 7th IEEE Latincom Conference, and the CQRM symposium Co-Chair in the IEEE ICC 2015 \& 2016 conferences. He is currently the General Co-Chair of the 22th IEEE CAMAD and the CQRM symposium Co-Chair in IEEE Globecom 2017, and the Chair of the IEEE ComSoc Technical Committee on Communication Systems Integration and Modeling (CSIM). 
\end{IEEEbiography}





\end{document}